\documentclass[a4paper,11pt]{article}
\pdfoutput=1 

\usepackage{jheppub}  
\usepackage[T1]{fontenc}  
\usepackage{amsmath}
\usepackage{slashed}
\usepackage{tikz}
\usepackage{graphicx}
\usepackage{pgfplots}
\usepackage{color}
\usepackage{rotating}
\usepackage{gensymb}
\usepackage{simplewick}
\usepackage{enumerate}
\usetikzlibrary{quotes,angles,snakes}
\usetikzlibrary{trees}
\usetikzlibrary{decorations.pathmorphing}
\usetikzlibrary{decorations.markings}
\usetikzlibrary{arrows,shapes,positioning}
\usetikzlibrary{decorations.markings}
\tikzstyle arrowstyle=[scale=1]
\tikzstyle directed=[postaction={decorate,decoration={markings,
    mark=at position .65 with {\arrow[arrowstyle]{stealth}}}}]
\tikzstyle reverse directed=[postaction={decorate,decoration={markings,
    mark=at position .65 with {\arrowreversed[arrowstyle]{stealth};}}}]
\DeclareMathOperator{\Tr}{Tr}
\newcommand{\RNum}[1]{\uppercase\expandafter{\romannumeral #1\relax}}

\newcommand{\beq}{\begin{equation}}
\newcommand{\eeq}{\end{equation}}
\newcommand{\bea}{\begin{eqnarray}}
\newcommand{\eea}{\end{eqnarray}}
\usepackage{ulem}

\usepackage{xcolor,colortbl}
\definecolor{Blue}{RGB}{140,165,195}
\definecolor{Purple}{RGB}{255,145,145}

\definecolor{rossoCP3}{cmyk}{0,.88,.77,.40}
\definecolor{graa}{rgb}{0.8,0.8,0.8}
\definecolor{blaa}{rgb}{0.2,0.2,0.6}
 
\definecolor{orange}{cmyk}{0,0.5,1,0}
\definecolor{rossoCP3}{cmyk}{0,.88,.77,.40}
\definecolor{graa}{rgb}{0.8,0.8,0.8}
\definecolor{blaa}{rgb}{0.2,0.2,0.6}
\hypersetup{
	colorlinks, 
	bookmarksopen, 
	bookmarksnumbered,
	citecolor=blaa, 		
	linkcolor=rossoCP3,	
	urlcolor=rossoCP3,			
}

\begin{document} 

\title{\boldmath \textcolor{rossoCP3} {Phase structure of complete asymptotically free SU($N_c$) theories with quarks and scalar quarks 
}}
 
\author[a]{F.F.~Hansen,}
\author[a]{T.~Janowski,}
\author[a]{K.~Lang{\ae}ble,}
\author[b]{R.B.~Mann,}
\author[a,c,d]{F.~Sannino,}
\author[e]{T.G.~Steele,}
\author[a,b]{Z.W.~Wang\,}

\affiliation[a]{$\rm{CP}^3$-Origins, University of Southern Denmark,\\Campusvej 55
5230 Odense M, Denmark}
\affiliation[b]{Department of Physics, University of Waterloo,\\Waterloo, ON, N2L 3G1, Canada}
\affiliation[c]{Theoretical Physics Department, CERN, Geneva, Switzerland}
\affiliation[d]{Danish IAS, University of Southern Denmark, Denmark}
\affiliation[e]{Department of Physics \& Engineering Physics, University of Saskatchewan,\\Saskatoon, SK, S7N 5E2, Canada}

\emailAdd{ffhansen@cp3.sdu.dk}
\emailAdd{janowski@cp3.sdu.dk}
\emailAdd{langaeble@cp3.sdu.dk}
\emailAdd{rbmann@uwaterloo.ca}
\emailAdd{sannino@cp3.sdu.dk}
\emailAdd{tom.steele@usask.ca}
\emailAdd{wang@cp3.sdu.dk}

\abstract{ We determine the phase diagram  of complete asymptotically free SU($N_c$) gauge theories featuring $N_s$ complex scalars and  $N_f$ Dirac quarks transforming according to the fundamental representation of the gauge group. The analysis is performed at the maximum known order in perturbation theory. We unveil a very rich dynamics and associated phase structure. Intriguingly we discover  that the complete asymptotically free conditions guarantee that the infrared dynamics displays long distance conformality, and in a regime when perturbation theory is applicable.  We conclude our analysis by determining the quantum corrected potential of the theory and summarising the possible patterns of radiative symmetry breaking. These theories are of potential phenomenological interest as either  elementary or composite ultraviolet finite extensions of the Standard Model. \\[5mm]
	{\it Preprint: CERN-TH-2017-133 \& CP3-Origins-2017-022 DNRF90}
 }

\maketitle
\flushbottom

\section{Introduction}
\label{sec:intro}

Gauge theories featuring gauge, scalar and fermion degrees of freedom constitute the back-bone of the Standard Model of particle interactions. It is therefore important to unveil their perturbative and non-perturbative dynamics. 

Furthermore according to their ultraviolet properties these theories can be classified into fundamental and effective low energy descriptions. Fundamental theories are, according to Wilson, the ones featuring in the UV non-interacting (free) \cite{Gross:1973ju,Gross:1973id,Gross:1974cs,Politzer:1973fx} or interacting (safe)  \cite{Pica:2010xq,Litim:2014uca} fixed points.  If multiple couplings are present one can have complete asymptotic freedom (CAF) \cite{Gross:1973ju,Cheng:1973nv,Callaway:1988ya,Giudice:2014tma,Holdom:2014hla,Pica:2016krb,Molgaard:2016bqf,Gies:2016kkk,Einhorn:2017jbs}, or safety (CAS) \cite{Litim:2014uca,Litim:2015iea,Esbensen:2015cjw}, or mixed possibilities \cite{Esbensen:2015cjw,Molgaard:2016bqf,Pelaggi:2017wzr}.   Exact non-perturbative results on the possible asymptotically (un)safe nature of supersymmetric gauge theories were investigated in \cite{Intriligator:2015xxa}  impacting the very existence of time-honoured super grand unified theories \cite{Bajc:2016efj}. The existence of controllable non-supersymmetric theories with interacting UV fixed points led to the recent discovery  \cite{Abel:2017ujy} that the addition of positive mass-squared terms leads to calculable radiative symmetry breaking in the IR, a phenomenon akin to the radiative symmetry breaking that occurs in the Supersymmetric Standard Model \cite{Ibanez:1982fr}. We will not consider gravitational corrections which, however, are the subject of interesting related work \cite{Eichhorn:2017eht,Christiansen:2017qca,Christiansen:2017gtg}. 

Here we focus our attention on the dynamics of $SU(N_c)$ gauge theories with $N_s$ complex scalars and  $N_f$ Dirac quarks transforming in the fundamental representation of the gauge group. Surprisingly despite the in depth study within the supersymmetric context, mostly due to the remarkable work by Intriligator and Seiberg \cite{Intriligator:1995au}, very little is known about the non-supersymmetric version with only one complex species of scalar quarks in addition to the ordinary quarks. 

We therefore wish to partially close this gap by providing an in depth study of these theories within a perturbative RG analysis along with the study of the associated quantum effective potential in the fully calculable regime. We discover a very rich physics associated to the various possible phases in which the theories can be. 

The choice to study these theories stems from the past and recent interest in elementary  \cite{Gross:1973ju,Cheng:1973nv,Callaway:1988ya,Giudice:2014tma,Holdom:2014hla,Pica:2016krb,Molgaard:2016bqf,Einhorn:2017jbs} and composite extensions of the Standard Model featuring scalar quarks both in models of (super) bosonic Technicolor \cite{Kagan:1991ng,Dobrescu:1995gz,Kagan:1994qg,Altmannshofer:2015esa} as well as in models of composite Higgs dynamics \cite{Kaplan:1983fs,Kaplan:1983sm,Dugan:1984hq} embodying explicit realisations  \cite{Sannino:2016sfx,Cacciapaglia:2017cdi} of the partial composite mechanism for standard model mass generation \cite{Kaplan:1991dc}. The underlying realisations\footnote{A list of underlying fundamental theories for near conformal dynamics and composite Higgs theories, before considering fermion mass generation, can be found in \cite{Sannino:2009za,Cacciapaglia:2014uja}.} of these  composite extensions are dubbed {\it fundamental partial composite theories}  \cite{Sannino:2016sfx}. 
 
 For the theories at hand we first investigate the CAF conditions. We then examine the infrared dynamics of the unveiled CAF theories to the maximum known order in perturbation theory, allowing us to determine the perturbative phase diagram. Since theories with scalars can undergo a radiative symmetry breaking phenomenon because of the Coleman Weinberg (CW) mechanism \cite{Coleman:1973jx}, we carefully investigate this possibility here using the improved Gildener Weinberg (GW) approach \cite{gildener,Gildener:1975cj}.   We show that under certain conditions these theories feature, besides CAF, also large distance conformality. 
 
 We now lay out the structure of the paper.  In section 2, we introduce the theories, their beta functions and spell out the conditions for CAF. We move to show the emergence of controllable interacting infrared fixed points to the maximum known order in perturbation theory. We discover that the phase diagram is rich and that the CAF conditions lead also to infrared conformality, at least in some coupling direction. Spontaneous radiative symmetry breaking is analyzed in section 3. Here we pay special attention to the possible patterns of symmetry breaking in the scalar sector. The analysis is performed in steps, with the zeroth order corresponding to a tree-level analysis and the quantum corrections studied at the one-loop order. The presence of multiple couplings leads to different limits in the parameter space of the theory that can affect the radiative breaking scenarios.  We conclude in section 4 and add a number of appendices containing further technical details.

\section{Ultraviolet and infrared properties of the theory}
\label{UV-IR}
 
In this work, we consider an $SU(N_c)$ gauge theory involving $N_s$ complex scalars $S$ and $N_f$ vector-like fermions $Q$ in the fundamental representation. The Lagrangian is
\begin{equation}
\mathcal L=-\frac{1}{2}\Tr F^{\mu\nu}F_{\mu\nu}+\Tr\left(\bar{Q}i\slashed{D}Q\right)+\Tr\left(D_\mu S^{\dagger}D^\mu S\right)-v \left(\Tr S^{\dagger}S\right)^2
-u \Tr\left(S^{\dagger}S\right)^2\,,
\end{equation}
where $F_{\mu\nu} = F_{\mu\nu}^a t^a\,\left(a=1\cdots,\,N_c^2-1\right)$ is the field strength tensor and $t^a$ are the SU($N_c$) generators in the fundamental representation satisfying $\Tr\left( t^at^b \right) = (1/2)\delta^{ab}$. In our notation, the fermion fields $Q$ and the scalar fields $S$ are rectangular matrices with dimensions $N_f \times N_c$ and $N_s \times N_c$ respectively. 

Note that for specific colour choices, with  $N_c \leq 4$ and $N_s \leq 4$,  additional renormalizable terms in the Lagrangian appear. Here for  example for $N_c=N_s$ we can construct the operator $\mathrm{det} S$. Similarly, for $N_c=4$ and $N_s=2$ or $N_c=2$ and $N_s=4$ we have terms of the form $\varepsilon_{ab}\varepsilon_{cd}\varepsilon^{ijkl}S^a_iS^b_jS^c_kS^d_l$. Furthermore for $N_c=3$ a Yukawa term can be written involving one scalar and two quarks. Additional terms of these types would give additional contributions to the beta functions, which are not considered in this work.

\begin{table}[ht]
\centering
\begin{tabular}{| c | c | c c c |}
  \hline 
  Fields & Gauge Symmetries &   \multicolumn{3}{c |}{Global Symmetries} \\  \hline

   & $SU(N_c)$ & $SU(N_f)_L$ & $SU(N_f)_R$ & $U(N_s)$ \\ 
\hline \hline
\rowcolor{Blue!40} 
  $Q$ & $\Box$ & $\Box$ & 1 & 1 \\ 
\rowcolor{Blue!40}
  $\tilde{Q}$ & $\overline{\Box}$ & 1 & $\overline{\Box}$ & 1 \\ \hline \hline
\rowcolor{Purple!40}
  $S$ & $\Box$ & 1 & 1 & $\Box$ \\
  \hline
\end{tabular}
\caption{Matter field content of the theory including the quantum symmetry group.  Fermion fields are presented in the left-handed spinor convention.}
\end{table}

\subsection{UV Behaviour: Completely Asymptotically Free (CAF)}\label{CAF}
Since the theory has three marginal couplings we now investigate its ultraviolet behaviour and establish the conditions under which it can be 
  complete asymptotically free.  
We are interested in characterising the flow behaviour around the Gaussian fixed point, and therefore one can use one-loop expressions for the beta functions. 

 Using the rescaled couplings, i.e. $\alpha = g^2 /(4 \pi)^2, \lambda_1 = v /(4 \pi)^2, \lambda_2 = u /(4 \pi)^2$, the one-loop beta functions are
\begin{align}
		\beta_{\alpha} &=  -\frac13 \left(22N_c-4N_f-N_s\right)\alpha^2\nonumber\\
		\beta_{\lambda_1 } &= ~4 (N_c N_s+4) \lambda_1 ^2 +12\lambda_2^2 \nonumber\\
		&\phantom{{}={}}+\lambda_1 \bigg[ 8(N_c+N_s)\lambda_2 - \frac{6(N_c^2-1)}{N_c} \alpha \bigg]+ \frac{3 (N_c^2 + 2)}{4N_c^2} \alpha^2 \nonumber \\
		\beta_{\lambda_2} &= 
		4(N_c+N_s)\lambda_2^2+\lambda_2 \bigg[24\lambda_1  - \frac{6(N_c^2-1)}{N_c} \alpha\bigg]+\frac{3 (N_c^2 - 4)}{4 N_c}\alpha^2.
	\label{betafunctions}
\end{align}
We note that, to this order, the gauge beta function only depends on the gauge coupling, and it is of the form $\beta_\alpha = - B \alpha^2$. Requiring asymptotic freedom (AF), is equivalent to restricting the coefficient $B$ to be positive. The critical number of fermion flavors, $N_{f}^{*}$, for which asymptotic freedom is lost in the gauge coupling, $B = 0$, is $ N_{f}^{*}=(22 N_c - N_s)/4$.
%
%
The gauge coupling will be AF for theories with non-negative integer values $N_f <N_{f}^{*}$. We thus have an upper bound on $N_s$, i.e. $N_s \leq 22N_c$.

The two beta functions for the quartic couplings depend on all three couplings but only on $N_c$ and $N_s$. 

For CAF to exist we first  need to find solutions to the following fixed flow relation  \begin{equation}
(\beta_{\alpha} , \beta_{\lambda_1}, \beta_{\lambda_2}) = c ( \alpha, \lambda_1, \lambda_2) \, ,
\label{eq:fixedflow}
\end{equation}
for an arbitrary non-zero constant $c$. 

We will now outline the criteria under which solutions exist and count the number of solutions.
 The fixed flow solution for $\alpha > 0$ will necessarily satisfy $c = - B\alpha$, which can be substituted into the remaining equations. 
Since $\beta_{\lambda_2}$ depends linearly on $\lambda_1$, we can easily solve for $\lambda_1 (\lambda_2, \alpha)$ and substitute this into the equation for $\beta_{\lambda_1}$.  The result is a quartic equation in $\lambda_2$. 
The coefficients of this  equation depend on $N_c, N_s, B$. Introducing the quantity $N_x=N_f^*-N_f$, we can express $B = 4 N_x / 3$. The number of fixed flow solutions corresponds to the number of real roots of the quartic polynomial, which can be calculated using the discriminant method. 
The full expression for the quartic polynomial and the expressions for classifying the nature of the roots can be found in Appendix~\ref{quartic}. 
For a fixed value of $N_x$, we find a region with two distinct real solutions (Region I) and a smaller region with four distinct real solutions (Region II). 
The upper border of the Region II is shared with Region I. 

\begin{figure}[t]
	\centering
	\includegraphics[width=0.6\columnwidth]{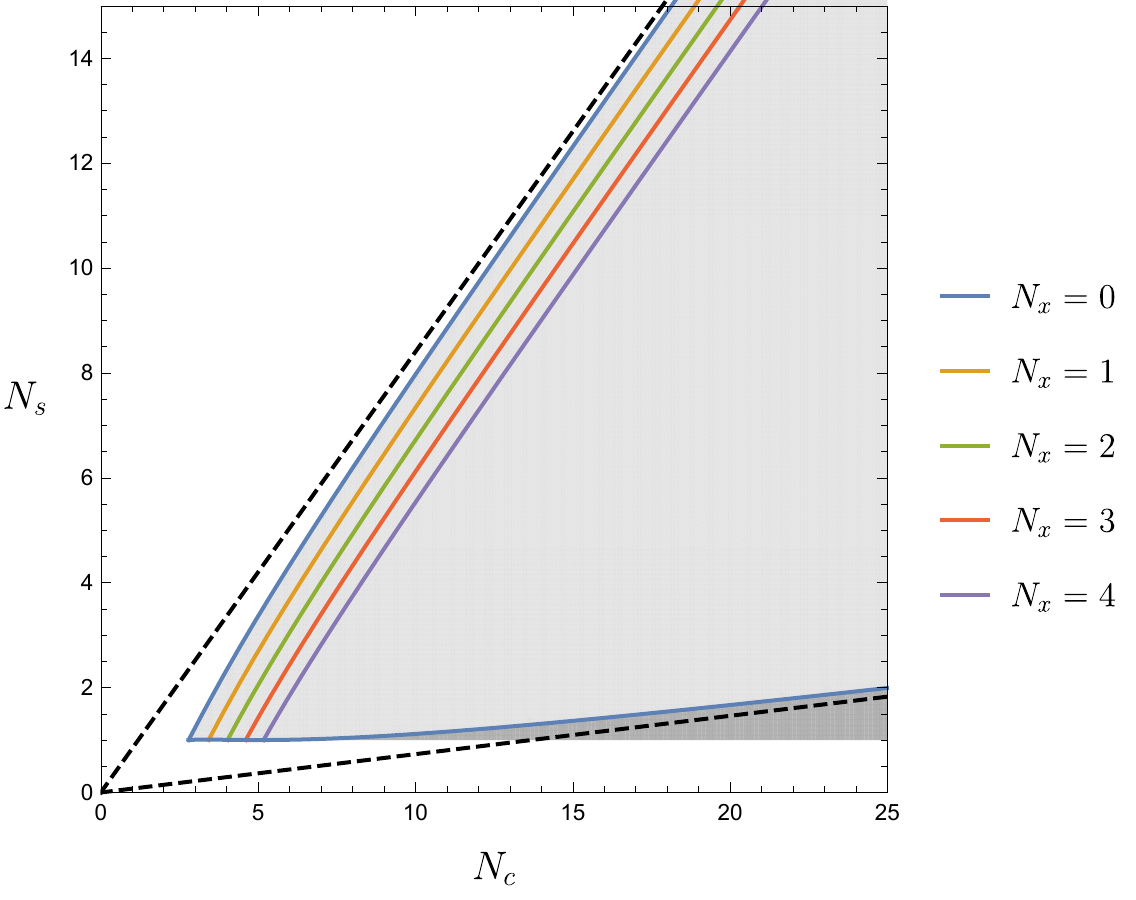}
		\caption{Regions in $N_c, N_s$ for constant $N_x$ for which the theory is CAF. The solid lines show the borders of the region with two fixed flow solutions (Region I) for $N_x \in \{0,1,2,3,4\}$ The blue boundary $N_x=0$ represents a limit and does not satisfy CAF. The light gray (gray) region marks the region where two (four) real distinct sets of pseudo fixed points exist. The dashed black lines are the asymptotic behavior of the borders of the two fixed flow solution region.}
	\label{Region2}
\end{figure}

In Fig.~\ref{Region2} we show how the borders of these regions change when varying the number of fermion flavours ($N_x \in \{0,1,2,3,4\}$). It should be noted, that for small values of $N_x$, the effect of increasing $N_x$ is to move the upper border of Region I downward, while the borders of Region II largely remains unchanged. 
Furthermore, we note that in the limit $N_x=0$ ($N_f=N_f^*$), Eq.~(\ref{eq:fixedflow} no longer describes fixed flows, since the one loop beta function for gauge coupling is vanishing. Nonetheless, solving the resulting quartic function corresponds to finding pseudo fixed points in the subsystem $\beta_{\lambda_1},\beta_{\lambda_2}$, i.e. fixed points in $\lambda_1, \lambda_2$, where $\alpha$ is treated as a constant.
The region with two distinct sets of pseudo fixed points is marked with light gray in Fig.~\ref{Region2}, while the region with four distinct sets of pseudo fixed points is marked with gray.  
From Fig.~\ref{Region2}, we note that for small values of $N_x$, the theories with fixed flow solutions also possess pseudo fixed points in the $\beta_{\lambda_1},\beta_{\lambda_2}$ subsystem. In Appendix~\ref{app:UVIR}, we elucidate the connection between the existence of fixed flow solutions and pseudo fixed points. 
From Fig.~\ref{Region2}, we see that for fixed values of $N_s \geq 2$ and $N_x >0$, there exists a lower bound of $N_c$ above which the theories have two fixed flow solutions and are CAF. For even higher values of $N_c$, two additional fixed flow solutions appear. Increasing either $N_x$ (i.e.~lowering $N_f$) or $N_s$ will push the lower bound on $N_c$ towards higher values, whereas the transition from two to four fixed flow solutions is only mildly dependent on $N_x$. 
In Table~\ref{flowwindow} the values of $N_c, N_s, N_f$ for theories with CAF are tabulated. 

Existence of fixed flows implies that at least the points along one direction of each of the flow lines flow out of the Gaussian fixed point. However, to complete the picture, we need information about the behaviour of the RG trajectories in the neighborhood of the fixed flow lines.
%
To investigate this, we parametrise the couplings  $\left(\alpha, \lambda_1, \lambda_2\right)$ using spherical coordinates 
\begin{equation}
  \alpha = r \sin\theta \cos\phi, \qquad
  \lambda_1 =r \sin\theta \sin\phi, \qquad
  \lambda_2= r \cos\theta\, ,
\end{equation}
and obtain expressions for the RG beta functions of $\left(r, \theta, \phi\right)$, i.e.~$(\beta_r,\,\beta_{\theta},\,\beta_{\phi})$.  

\begin{figure}[t]
	\centering
	\includegraphics[width=0.47\columnwidth]{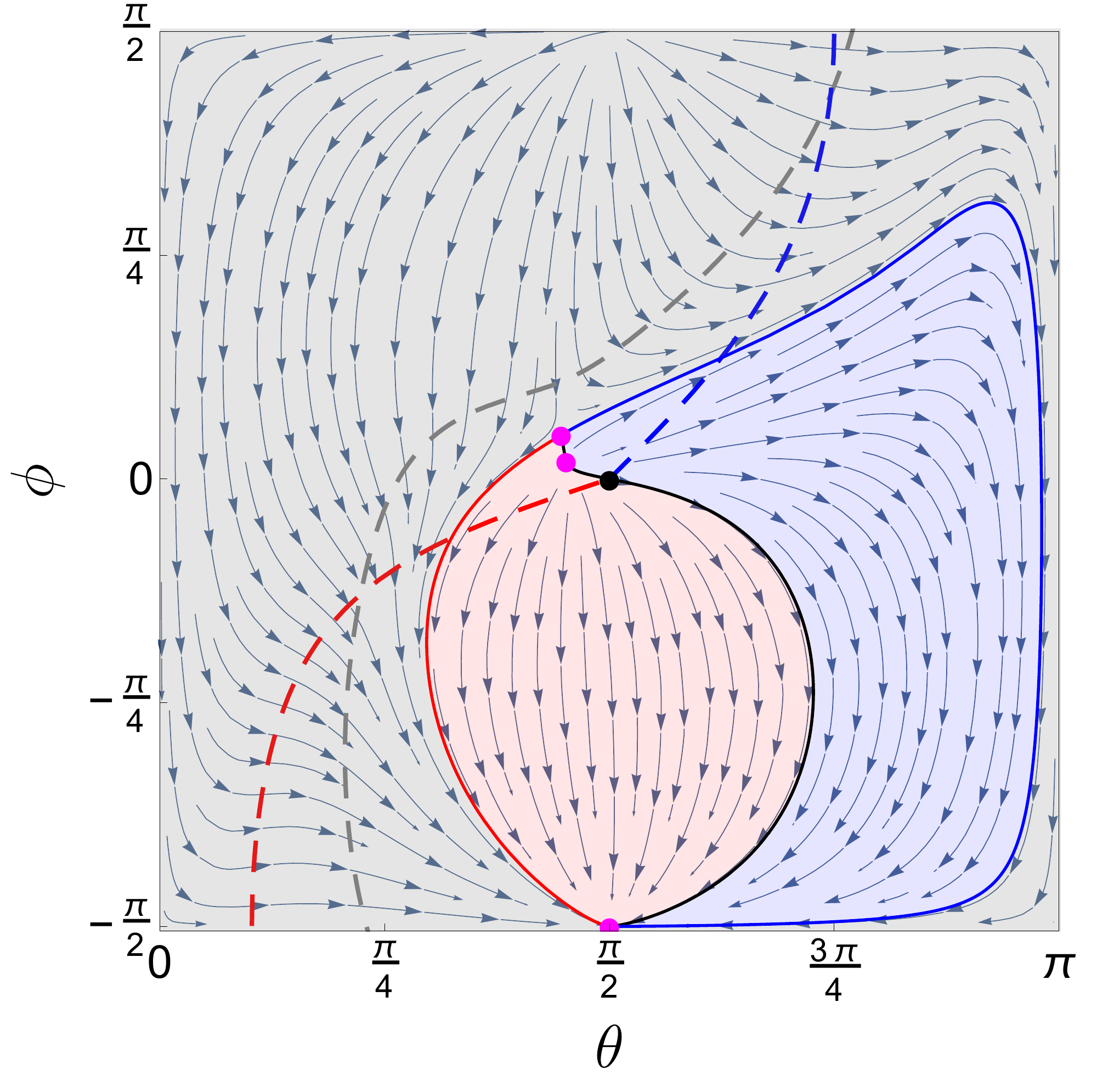}
	\hspace{0.01\columnwidth}
	\includegraphics[width=0.47\columnwidth]{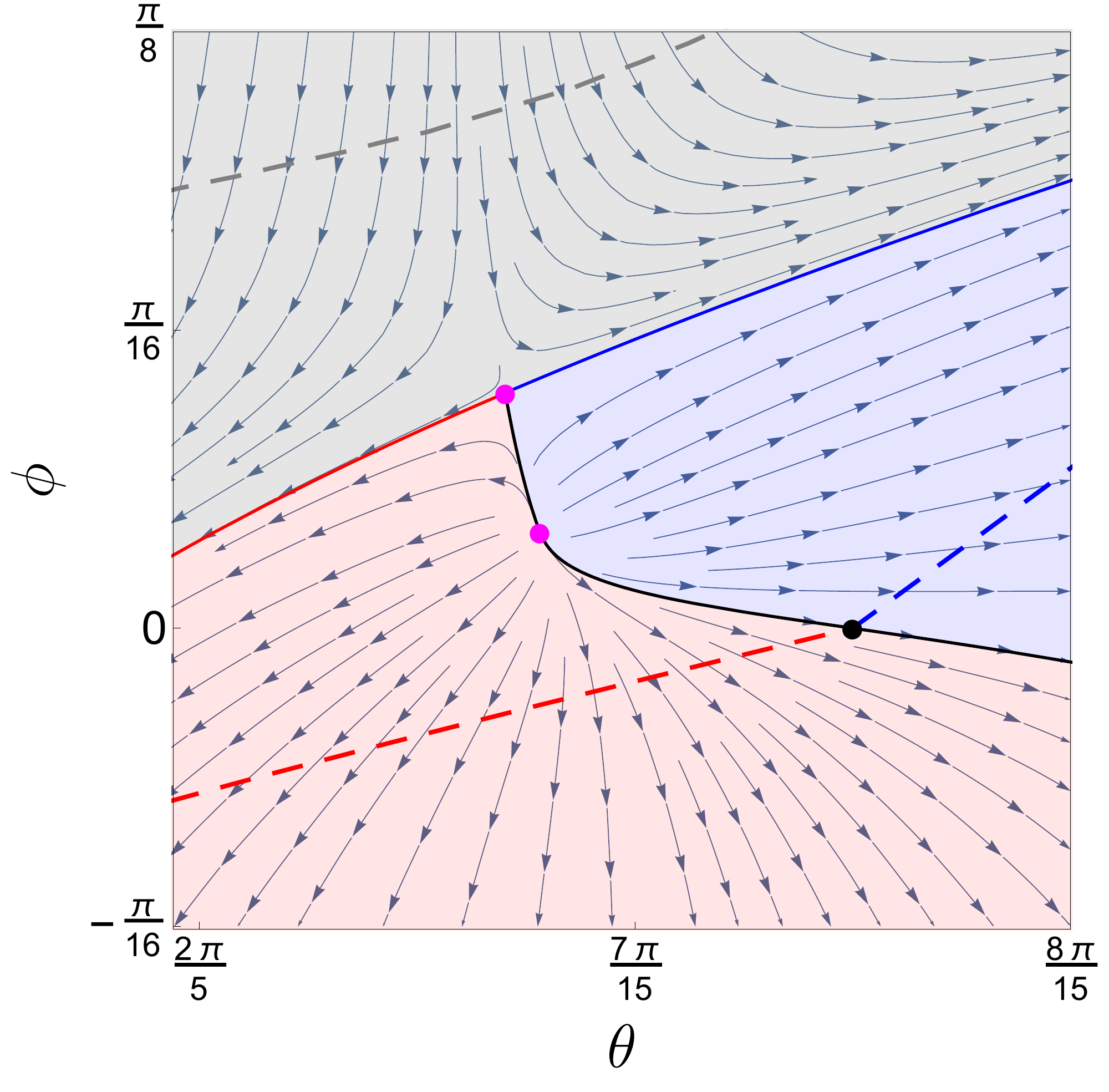}\vspace{0.01\columnwidth}
	\includegraphics[width=0.47\columnwidth]{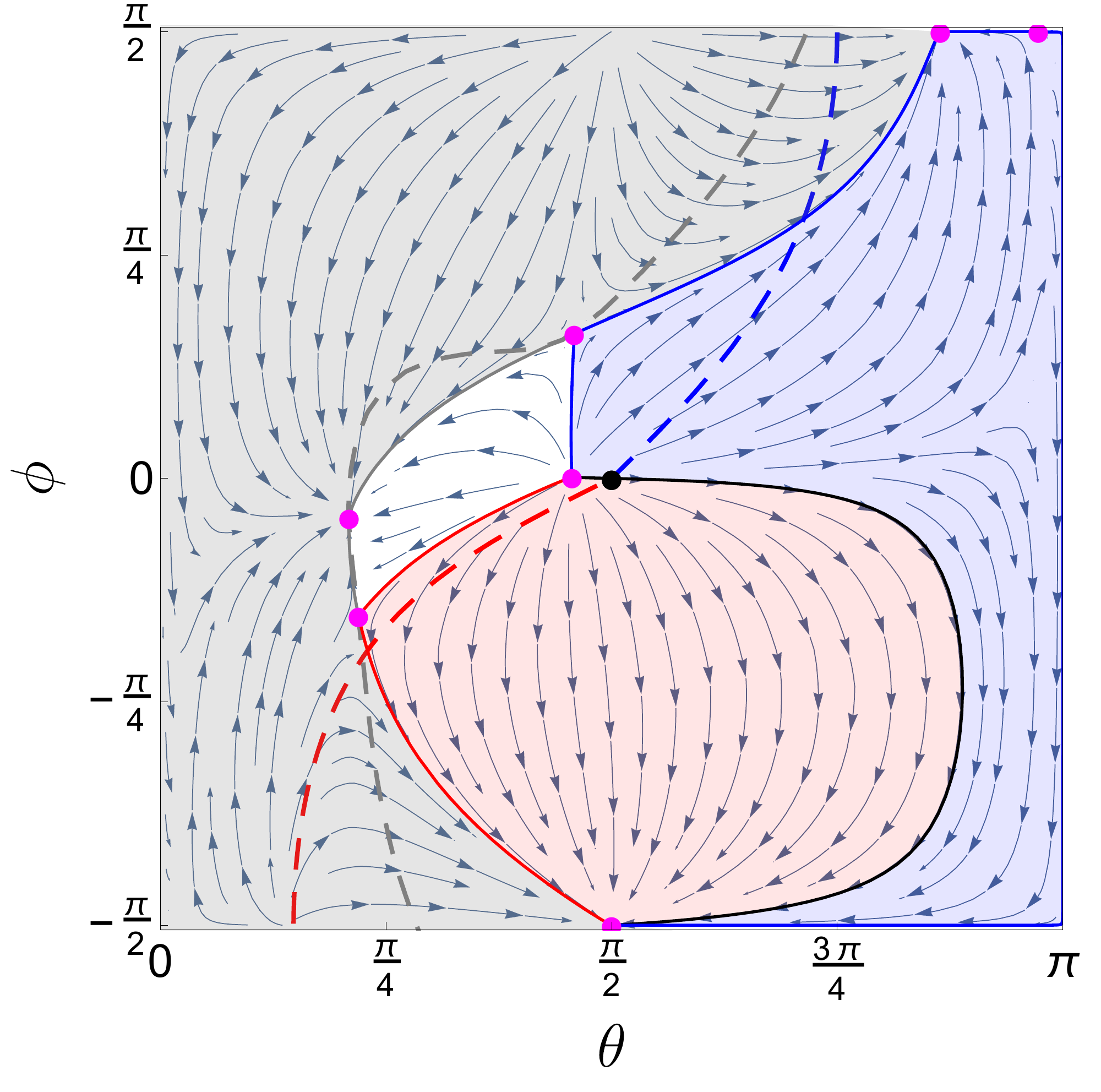}
	\hspace{0.01\columnwidth}
	\includegraphics[width=0.47\columnwidth]{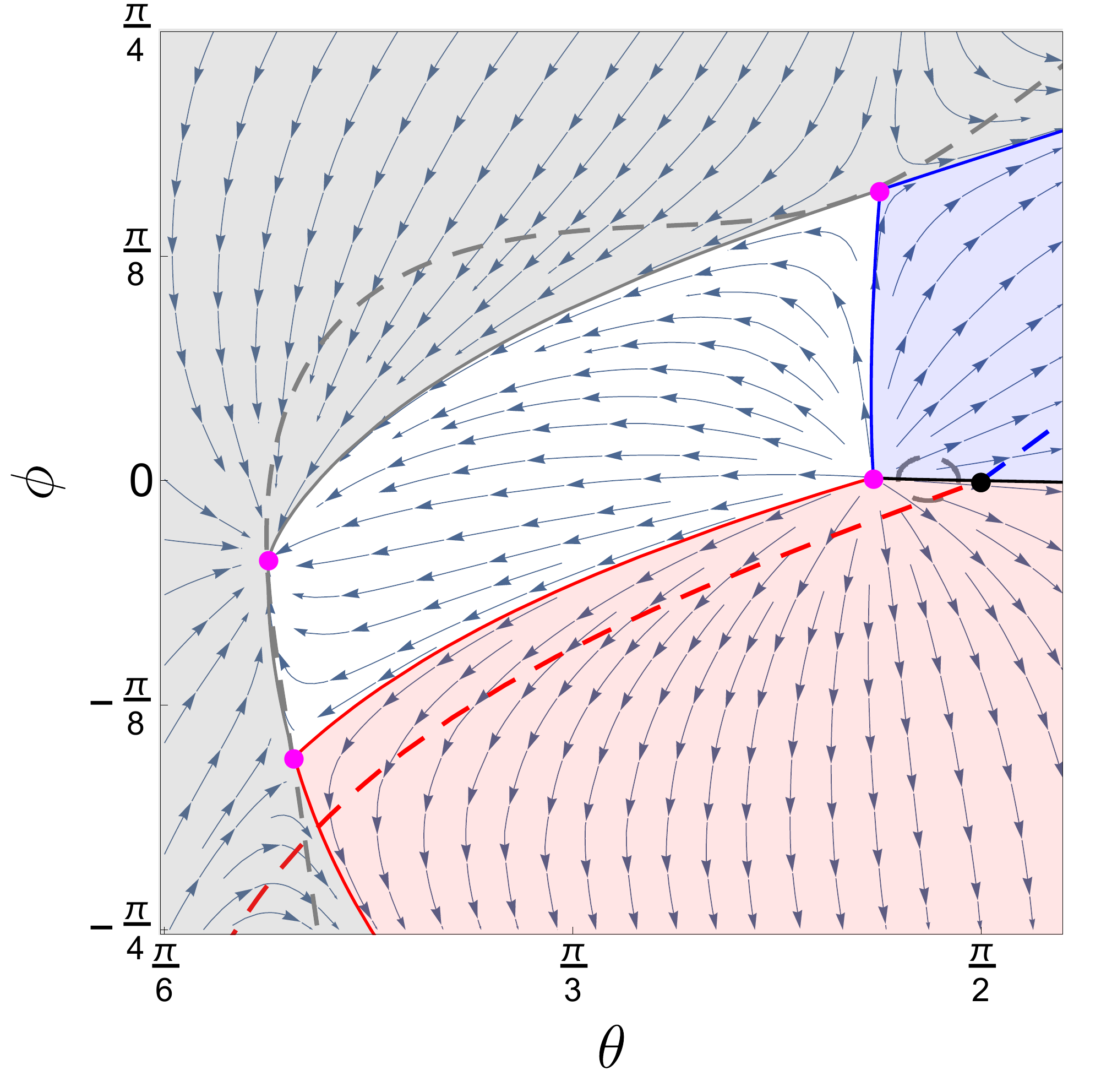}
	\caption{UV behavior around the fixed flow solutions in spherical coordinates $\left( \theta, \phi\right)$. Upper left: Full phase space in case of two fixed flow solutions. Upper right: Close-up at the region close to the two fixed points. Lower left: Full phase space in case of four fixed flow solutions. Lower right: Close-up at the region close to the four fixed points. Flows not connected to the UV fixed point are in grey regions. Flows  between fixed flows that do not cross the tree level symmetry breaking lines are in white regions; those that do cross are
in red or blue regions as discussed in Sec.~\ref{tree-analysis}.}
	\label{AngularPlot}
\end{figure}

At one loop, the beta functions, $\beta_{\theta},\,\beta_{\phi}$, depend only multiplicatively on the radial coordinate, and the direction $\left( \theta, \phi\right)$ therefore does not depend on $r$. Factoring out the radial coordinate $r$, we can examine the UV behavior in a reduced space of only two parameters. This is shown in Fig.~\ref{AngularPlot} for a theory in Region I (top panels) and a theory in Region II (bottom panels), with the arrows pointing from UV to IR.

In order for a trajectory to be connected to the Gaussian UV fixed point, the radial coordinate has to go to zero in the UV. Within the one-loop approximation, the change in $r$ is of the form $\beta_r = r^2\,f(\theta,\phi)$. The radial coordinate is thus only decreasing in regions of $\left( \theta, \phi\right)$ where $f(\theta,\phi)<0$. In Fig.~\ref{AngularPlot}, the transition ($f(\theta,\phi)=0$) is shown as dashed grey lines.  The regions with flows that cross this line, and are thus not connected to the Gaussian UV fixed point, are colored grey. Similarly we mark the regions with red and blue, where the flows cross the tree level symmetry breaking boundary lines (dashed red and blue lines), derived in Sec.~\ref{tree-analysis}.

We show the UV behavior for the full phase space in the case of two solutions in the upper left panel of Fig.~\ref{AngularPlot}, and a close-up of the region around the two fixed flow points in the upper right panel. We note that one point is completely repulsive, while the other has one repulsive direction and one attractive direction. For the mixed case, the repulsive direction separates the flows originating from the UV fixed point from the ones that are UV divergent. The attractive direction separates the two tree level symmetry breaking regions.
\begin{table}[ht]
	\centering
	\resizebox{\linewidth}{!}{%
		\begin{tabular}{| c | c | c | c | c | c | c | c | c |} 
			\hline
			& $N_c = 5$ & $N_c = 6$ & $N_c = 7$ & $N_c = 8$ & $N_c = 9$ & $N_c = 10$ & $N_c = 11$ & $N_c = 12$ \\ \hline
			$N_s = 2$   & $25-26$ & $29-32$ & $33-37$ & $37-43$ & $41-48$ & $44-54$ & $48-59$ & $52-65$ \\ \hline
			$N_s = 3$   & & $31-32$ & $35-37$ & $38-43$ & $42-48$ & $46-54$ & $50-59$ & $54-65$ \\ \hline
			$N_s = 4$   & & & $36-37$ & $40-42$ & $44-48$ & $48-53$ & $52-59$ & $55-64$ \\ \hline
			$N_s = 5$   & & & $37$ & $41-42$ & $45-48$ & $49-53$ & $53-59$ & $57-64$ \\ \hline
			$N_s = 6$   & & & & & $47$ & $51-53$ & $55-58$ & $59-64$ \\ \hline
			$N_s = 7$   & & & & & & $52-53$ & $56-58$ & $60-64$ \\ \hline
			$N_s = 8$   & & & & & & & $58$ & $62-63$ \\ \hline
			$N_s = 9$   & & & & & & & & $63$ \\ \hline
		\end{tabular}}
		\caption{Windows in $N_f$ for $N_s = \{2,9\}$, $N_c = \{5,12\}$ for which the theory is CAF, i.e. allow for fixed flow solutions to Eq.~(\ref{eq:fixedflow}. There exist two fixed flow solutions for every value of $N_f$ in each window. There are no solutions for $N_c = \{3,4\}$ and $N_s>1$.}  \label{flowwindow}
	\end{table}
In the lower two panels of Fig.~\ref{AngularPlot} we show the case of four fixed flow solutions. In this case, the two additional fixed points for the fixed flow open up a two-dimensional region where the flows are between fixed points without crossing the tree level symmetry breaking lines. This region is marked with white. One of the two additional fixed points for the fixed flows is fully attractive, while the other is with mixed properties. For the fully attractive one, only this exact relation of the three couplings is connected to the UV. This case therefore offers full predictability in the IR. This will be further discussed in Sec.~\ref{Phase}.

In the results discussed above, we factor out the $r$-dependence of the couplings, which is valid within the one-loop approximation of the beta functions. Therefore Fig.~\ref{AngularPlot} is only adequate for describing the behavior close to the Gaussian UV fixed point. 
Starting from a point where $r\ll1$, such that the beta functions are well approximated by the one-loop expressions, we trust the flows in the backward direction (towards higher energies) outside the grey regions, since $r$ is decreasing.  On the contrary, we cannot follow the flows too far forwards, since the approximation is getting worse ($r$ increasing). Still, seen from the UV perspective of the Gaussian fixed point, we expect the four-solution case to offer more possibilities to flow from the UV Gaussian fixed point to a possible IR fixed point because of the fully attractive fixed flow solution, which provides a region that seemingly does not cross the symmetry breaking lines.\\ 
  
\noindent Here we summarize our conclusions from restricting the theory to be CAF
:
\begin{itemize}

\item For fixed values of $N_s$ and $N_c$ within the regions of Fig.~\ref{Region2} there exists a window in $N_f$ (with upper endpoint given by $N_f^*$), for which the theories are CAF.
\item The size of the window depends on the position in the grey region of Fig.~\ref{Region2}. Close to the upper border (i.e.~$N_x=0$ line) of the light grey region, the size of the window is vanishing. This behaviour is evident from Table~\ref{flowwindow}, where we show the range of the window in $N_f$ for the lowest combinations of $N_s$ and $N_c$.
\item We find that the theories which are CAF, i.e. where UV attractive fixed flows exist, are a subset of the theories which have pseudo fixed points in the subsystem $\beta_{\lambda_1},\beta_{\lambda_2}$, i.e. fixed points in $\lambda_1, \lambda_2$, where $\alpha$ is treated as a constant. In App.~\ref{quartic} and \ref{app:UVIR}, we show that this statement is independent of the value of $\alpha$, as long as $\alpha>0$. In the next section, we will argue that the pseudo fixed points at higher loop order become physical fixed points. 
\item In Fig.~\ref{AngularPlot}, we illustrate the UV behavior in the vicinity of the Gaussian fixed point. In the two fixed flow solution case, there is one fully UV attractive direction and one with mixed properties. In the four fixed flow solution case, one of the two additional directions is fully UV repulsive while the other is mixed. 
\end{itemize}

\subsection {Long distance conformality}\label{IRbehaviour}
In the previous section, we found that the theories which are CAF in the UV, possess pseudo fixed points in the $\beta_{\lambda_1},\beta_{\lambda_2}$ subsystem  at one-loop for a fixed value of $\alpha$. In other words, if the gauge coupling has a fixed point ($\beta_\alpha = 0$), then we already know that there exist values $\lambda_1, \lambda_2$ satisfying $\beta_{\lambda_1} = 0,\beta_{\lambda_2} = 0$ at one loop in the quartic subsystem. Following the ordering from the Weyl-consistency conditions \cite{Antipin:2013pya,Jack:1990eb,Osborn:1989td}, we should treat the system at three loops in the gauge beta function together with the one-loop quartic beta functions. To keep the analysis in the main section light, we study here the two-loop gauge beta function together with the one-loop quartic beta functions, while the full result is presented in the appendix. At this order, the running of the gauge still decouples from the quartic couplings. We derive the gauge fixed point, and compute the accompanying fixed points values of the quartic couplings. We find that in most cases all three couplings at the fixed points are perturbatively small, and we therefore do not expect the three-loop contributions to the running of the gauge couplings to quantitatively change the preliminary findings of the this section. In fact, we find that each higher loop order contribution will be suppressed by a factor of $N_x/N_c$. This on the other hand requires us to restrict our IR analysis to theories close to losing AF, such that $N_x \ll N_c$, to achieve perturbative control of the loop expansion. 
In the generalized Veneziano-limit, where $N_c,N_s,N_f \to \infty$ in such a way that all ratios are kept constant, the loop suppression can be arbitrarily small, whereas for finite values of $N_f$, $N_s$ and $N_c$ the smallest value for $N_x$ is $N_x = 1/4$. Further details are found in Appendix~\ref{app:three-loop}.

If we write the two-loop gauge beta function as $\beta_{\alpha}=-B \alpha^2 + C \alpha^3$, then the non trivial fixed point occurs for $\alpha^* = B/C$. For $B>0$ and $C>0$, this is an IR fixed point. The two loop coefficient for the gauge beta function takes the form
\beq
C =\frac{8}{3} N_c N_s - \frac{68 N_c^2}{3} - \frac{2 N_f}{N_c} + \frac{26  N_c N_f}{3} - \frac{2 N_s}{N_c} 
\eeq
The critical number of fermion flavors for this coefficient to be positive is
\beq
\bar{N}_f =\frac{34 N_c^3-4 N_c^2 N_s+3 N_s}{13 N_c^2-3}\, 
\eeq
and since in our case, where $N_c>2$, we always have that $N_f^*>\bar{N}_f$, the functions $\bar{N}_f$ and $N_f^*$ define the window of existence for the infrared fixed point in the gauge coupling. From requiring the theory to be CAF, we already know from the UV analysis, that if the gauge beta function has a non-trivial fixed point, then so does the quartic coupling subsystem. Therefore, the window for the existence of IR fixed points is the grey region in Fig.~\ref{Region2} for $N_f$ within $N_f\in [\bar{N}_f,N_f^*]$. The result is summarized in Table~\ref{fpwindow}, where the fixed points are determined numerically, and non-perturbative fixed points ($\alpha, \lambda_i > 1$) have been discarded.
 {\it Comparing table~\ref{flowwindow} with table~\ref{fpwindow}, we conclude that the CAF condition is stronger than the condition for the existence of IR fixed points, meaning that any CAF theory  displays long distance conformality. }
 
In order for the results not to be significantly altered by higher order contributions, we need to show that these higher loop-contributions are suppressed. This is shown in the appendix. Here it is sufficient to say that a Banks-Zaks-like analysis is possible. 
%
\begin{table}[ht]
	\centering
	\resizebox{\linewidth}{!}{%
		\begin{tabular}{| c | c | c | c | c | c | c | c | c | c |} 
			\hline
			 & $N_c = 4$ & $N_c = 5$ & $N_c = 6$ & $N_c = 7$ & $N_c = 8$ & $N_c = 9$ & $N_c = 10$ & $N_c = 11$ & $N_c = 12$ \\ \hline
			$N_s = 2$  & $11-21$ & $14-26$ & $16-32$ & $19-37$ & $21-43$ & $24-48$ & $27-54$ & $29-59$ & $32-65$ \\ \hline
			$N_s = 3$  & & $13-26$ & $16-32$ & $18-37$ & $21-43$ & $24-48$ & $26-54$ & $29-59$ & $31-65$ \\ \hline
			$N_s = 4$  & & & $16-31$ & $18-37$ & $21-42$ & $23-48$ & $26-53$ & $29-59$ & $31-64$ \\ \hline
			$N_s = 5$  & & & & $18-37$ & $20-42$ & $23-48$ & $26-53$ & $28-59$ & $31-64$ \\ \hline
			$N_s = 6$  & & & & & $20-42$ & $23-47$ & $25-53$ & $28-58$ & $31-64$ \\ \hline
			$N_s = 7$  & & & & & & $22-47$ & $25-53$ & $28-58$ & $30-64$ \\ \hline
			$N_s = 8$  & & & & & & & & $27-58$ & $30-63$ \\ \hline
			$N_s = 9$  & & & & & & & & & $30-63$ \\ \hline
		\end{tabular}}
		\caption{Window in $N_f$ that allow for IR fixed points with perturbative couplings, i.e. $\alpha, \lambda_i < 1$ for $N_s = \{2,9\}$, $N_c = \{4,12\}$. There are no solutions for $N_c = 3$ for $N_s >1$.}
		\label{fpwindow}
	\end{table}

\begin{figure}[t]
	\centering
	\includegraphics[width=0.47\columnwidth]{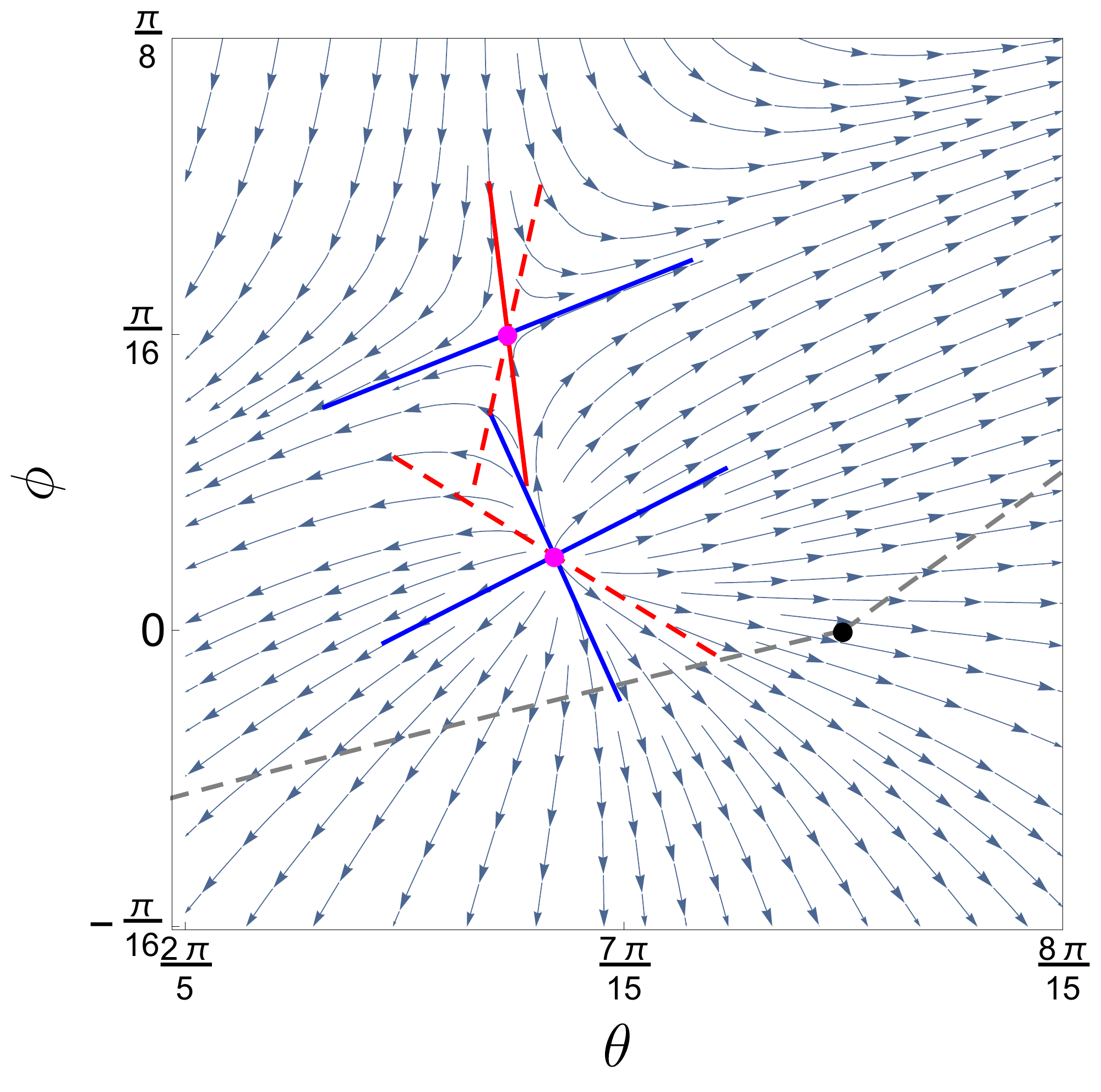}
	\hspace{0.01\columnwidth}
	\includegraphics[width=0.47\columnwidth]{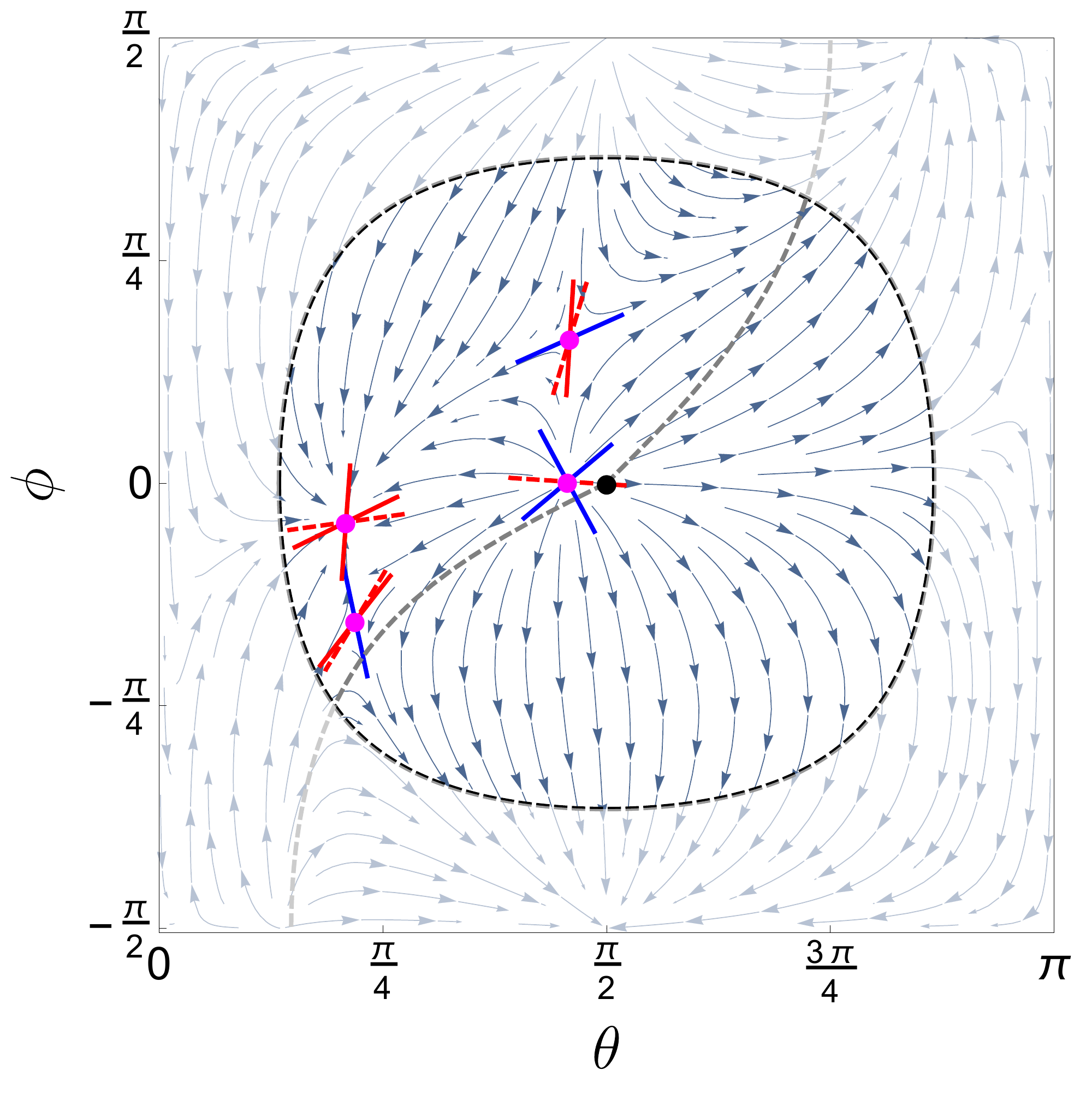}
	\caption{IR flow behavior around the IR fixed points in spherical coordinates $\left( \theta, \phi\right)$ with $\alpha$ kept fixed at the IR fixed point, $\alpha_* = B/C$. Left: Close-up of the case with two IR fixed points. Right: Full phase space in the case of four IR fixed points. Each fixed point (magenta dot), has its eigendirections superimposed. Color coding: Red is IR attractive, Blue IR repulsive. The red dashed lines are the projections of the third eigen-directions (IR attractive) of the fixed point in $\left(r,  \theta, \phi\right)$ onto the $\left( \theta, \phi\right)$-subspace. The shaded white region marks the region in $\left( \theta, \phi\right)$, where the quartic couplings become comparable to the gauge coupling at the fixed point, $\lambda_1^2 +\lambda_2^2 = 5 \alpha_*^2$, and higher order corrections are expected. }
	\label{IRangularPlot}
\end{figure}

In the following, we will characterize the IR fixed points.   
To find eigendirections of the IR fixed points, we need to study the following matrix:

\begin{eqnarray}
M &=& \left.\left( 
\begin{array}{ccc}
\frac{\partial \beta_\alpha}{\partial \alpha} & \frac{\partial \beta_\alpha}{\partial \lambda_1} & \frac{\partial \beta_\alpha}{\partial \lambda_2} \\
\frac{\partial \beta_{\lambda_1}}{\partial \alpha} & \frac{\partial \beta_{\lambda_1}}{\partial \lambda_1} & \frac{\partial \beta_{\lambda_1}}{\partial \lambda_2} \\
\frac{\partial \beta_{\lambda_2}}{\partial \alpha} & \frac{\partial \beta_{\lambda_2}}{\partial \lambda_1} & \frac{\partial \beta_{\lambda_2}}{\partial \lambda_2}
\end{array}
\right)\right|_{\alpha=\alpha^{*}, \lambda_1 = \lambda_1^*, \lambda_2 = \lambda_2^*} \,,
\end{eqnarray}
where $\left(\alpha^{*}, \lambda_1^*, \lambda_2^*\right)$ corresponds to the coupling solutions at the IR fixed points. In the convention that RG flow runs from UV to IR, the positive (negative) eigenvalues represent IR attractive (repulsive) directions. 
In the region, where we have perturbative control of our IR fixed points, they inherit their characteristics from the corresponding fixed flow solutions. The third eigendirection, which is not part of the fixed flow picture (Fig.~\ref{AngularPlot}), is dominated by the gauge coupling, and is always IR attractive. Within the region with two IR fixed points, we have one, which has two repulsive eigendirections, while the other fixed point has one repulsive and one attractive eigendirection. Combined with the third eigendirection, we can conclude that the one with two repulsive directions is only connected to the UV along a single trajectory, while the other one is connected through a one parameter family of trajectories. 
These trajectories all originate from the fully repulsive fixed flow direction. There is also a trajectory connecting the two IR fixed points. In Fig.~\ref{IRangularPlot}, we show the flow behavior around the IR fixed points on the plane of constant $\alpha=\alpha^*$ in spherical coordinates. This allows us to see the similarities with the UV picture (Fig.~\ref{AngularPlot}).
In the case with four IR fixed points, we have additionally one fixed point with mixed properties and one which is fully IR attractive. Unlike from the previous case, the fully IR attractive fixed point has a three dimensional basin of attraction, implying IR conformal stability in all directions. In Fig.~\ref{IRangularPlot}, this is illustrated by a two dimensional region, since $\alpha$ is kept fixed. In table~\ref{tab:fp}, we provide a summary of the IR fixed points.
\begin{figure}[t]
	\centering
	\includegraphics[width=0.9\columnwidth]{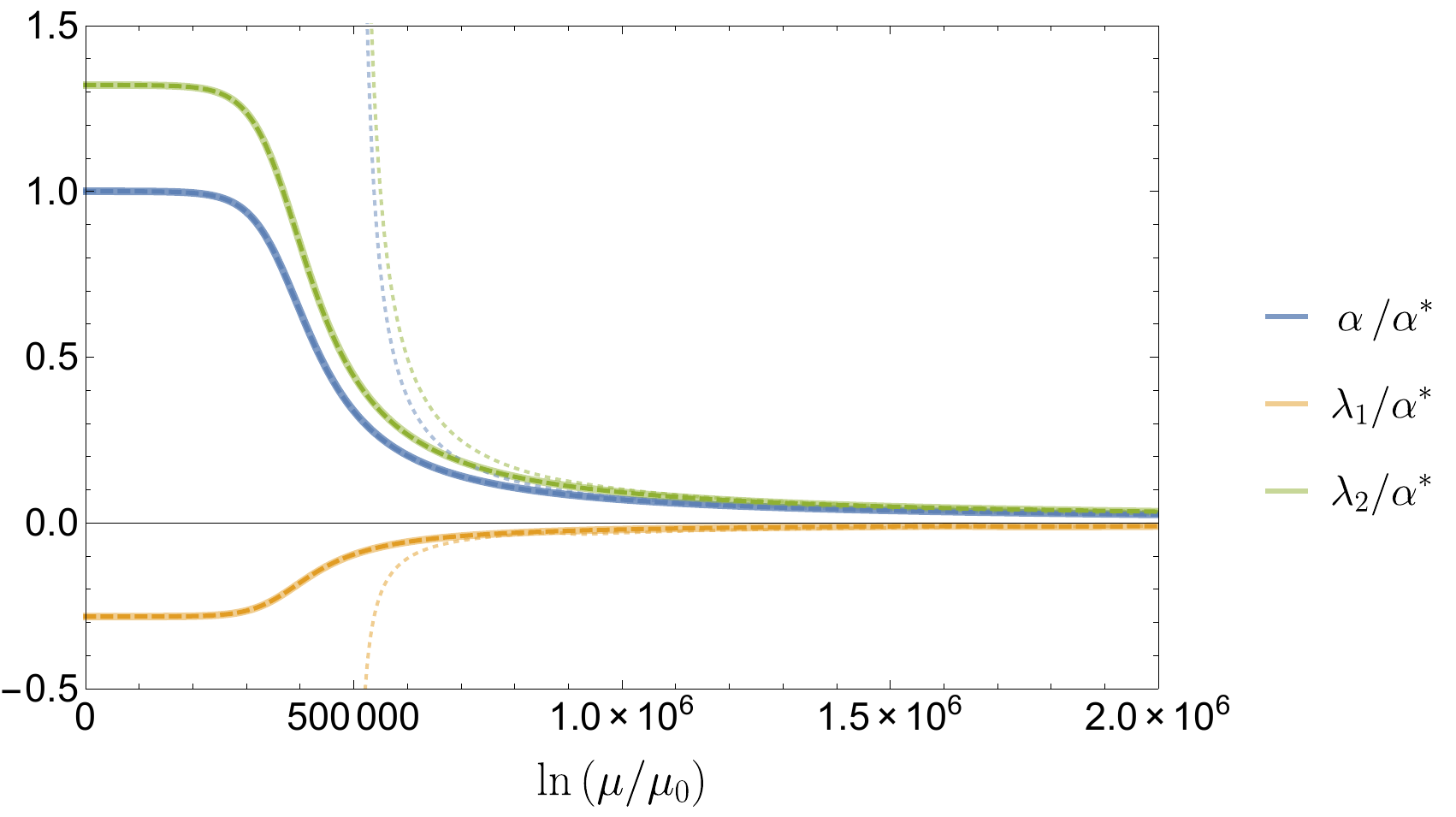}
	\caption{Renormalisation group running of all couplings from the fully IR attractive fixed point to the fully UV repulsive fixed flow. All couplings are normalised in units of $\alpha^*$. We use the one-loop beta functions for the quartic couplings together with the one-loop (dotted), two-loop (dashed) and three-loop (solid) gauge beta function.}
	\label{fig:RGflow}
\end{figure}
	\begin{table}
		\centering
		\begin{tabular}{c|ccc}
			\hline
			\hline 
			\multicolumn{1}{c}{Fixed Point } & \multicolumn{3}{c}{Eigenvalue} \\
			\hline FP$_1$ & + & -- & --\\ 
			FP$_2$  & + & -- & + \\ 
			FP$_3$ & + & + & + \\ 
			FP$_4$  & + & + & -- \\ 
			\hline
			\hline 
		\end{tabular}
		\caption{\label{tab:fp} Summary of IR fixed points and the number of their relevant and irrelevant eigenvalues. }
	\end{table}
%


We  learn that CAF theories with spin zero and spin half quarks also feature IR interacting fixed points.   In Appendix~\ref{app:UVIR}, we explore the connection in detail. In short, we can show that for fixed flows to exist with both $\lambda_1$ and $\lambda_2$ positive, there must  be pseudo fixed points in the quartic subsystem of beta functions. These pseudo fixed points become physical fixed points since the gauge coupling has an IR fixed point as well ($C>0$ in the region with pseudo fixed points). The case where both $\lambda_1$ and $\lambda_2$ are negative can be discarded by demanding a stable scalar potential. The cases with one of the quartic coupling negative, could be realised without the existence of pseudo fixed points. However, for the particular $N_c$ and $N_s$ dependence of the coefficients for this model, this does not occur.     

\subsection{Phase Diagram}\label{Phase}
From the previous section, we know there exist two kinds of phase structures; one with two IR fixed points and another with four IR fixed points. On one hand, the four IR fixed points case requires quite large number of colours $N_c$ and flavours $N_f$ even for small values of $N_s$. On the other hand, this scenario provides  a fully IR predictive case. In other words, this group of theories possesses a fully UV repulsive fixed flow, for which all relations among couplings are fixed, and we can thus fully determine the IR fate of the theory at the highest known order in perturbation theory. We will in the following do this for the minimal choice of colors, i.e. $N_c=26$, which in order to satisfy the CAF conditions requires $N_s = 2$ and $N_f = 138$. 
Afterwards, we will focus our attention on the general phases of the two types of phase structures. 

In Fig.~\ref{fig:RGflow}, for our particular choice of $N_c$, $N_s$ and $N_f$, we show the running of the couplings from the UV (with coupling ratios fixed by the fully repulsive fixed flow) towards the IR. We show the result from using both the one-, two- and three-loop gauge beta function together with the one-loop beta function for the quartic couplings. 
It is evident, that both in the two-loop and three-loop gauge case, the IR fate of the theory is long distance conformality.  
%
%
\begin{figure}[t]
	\centering
	\includegraphics[width=0.47\columnwidth]{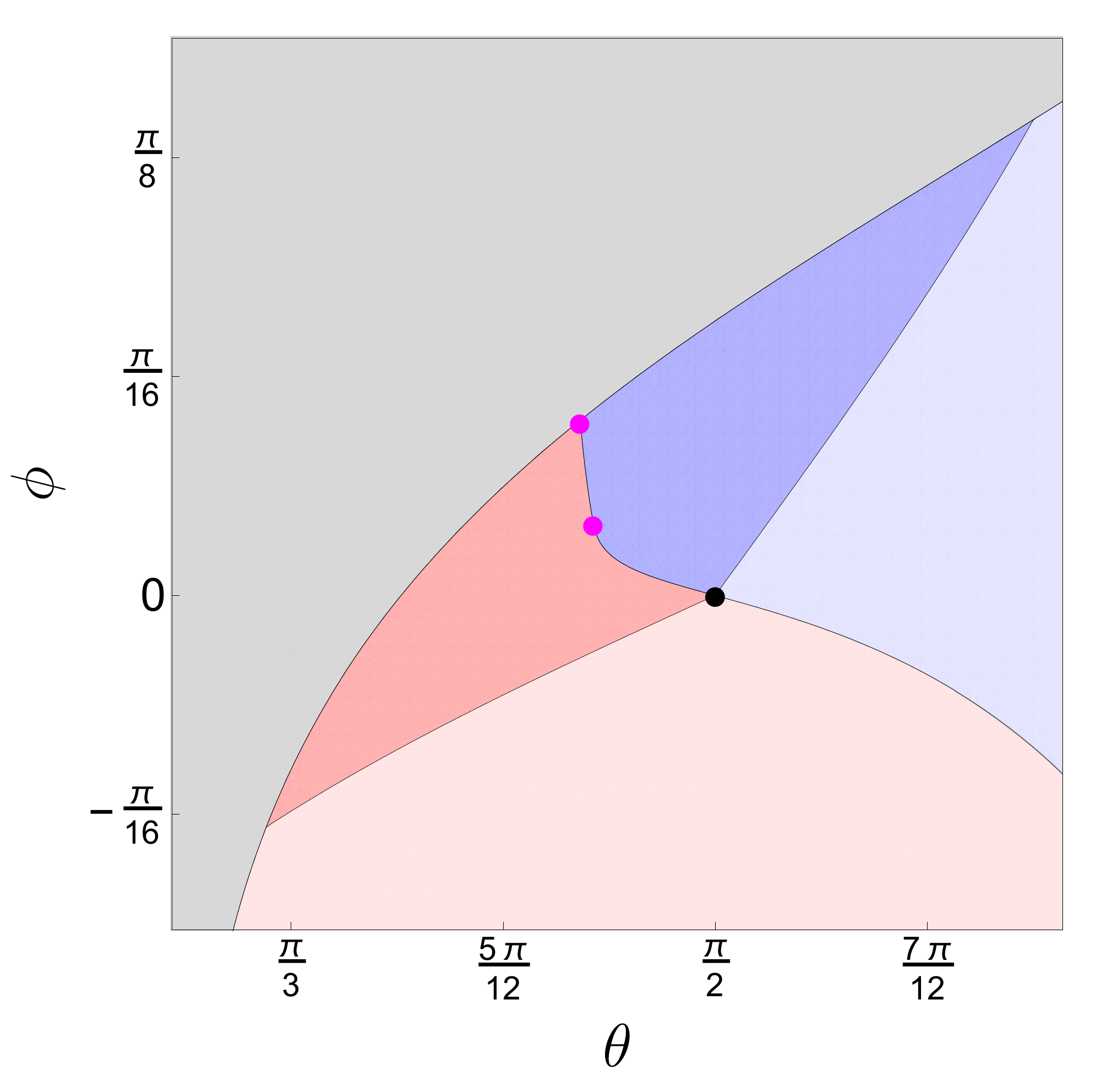}
	\hspace{0.01\columnwidth}
	\includegraphics[width=0.47\columnwidth]{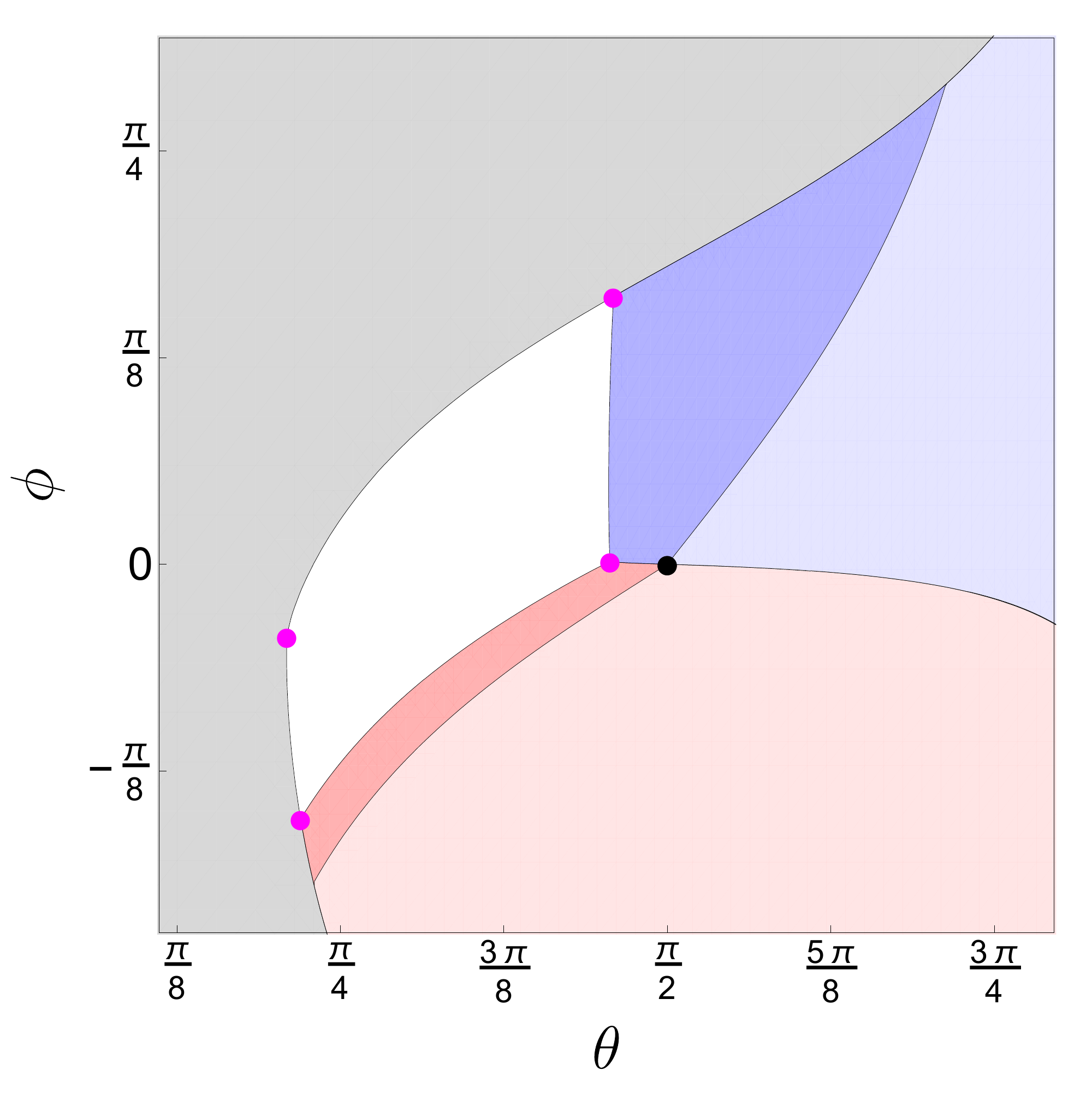}
	\caption{Phase diagram for the case of two (four) fixed flows are shown in the left (right) panel in spherical coordinates $\theta$ and $\phi$. The initial radial coordinate, $r$, is chosen close to the Gaussian fixed point, i.e. $r \ll 1$. The phases are then determined by analyzing the numerical solutions to the beta functions at three-loop order in the gauge coupling, together with the one-loop for the quartic couplings. The diagrams illustrate three IR phases connected to the Gaussian UV fixed point. Long distance conformality  (white), two different spontaneous symmetry breaking patterns (blue and red). 
	Directions with trajectories not originating from the Gaussian UV fixed point are colored gray. The light red and light blue regions correspond to initial conditions with unbounded tree level scalar potentials.}
	\label{PhaseDiagram}
\end{figure}

Similarly, we can solve the differential equations for each direction out of the Gaussian UV fixed point, and determine the IR fate anticipating and using the symmetry breaking conditions derived in Sec.~\ref{symmetrybreaking}. In this way, we distinguish between three IR phases connected to the Gaussian UV fixed point. Long distance conformality  (white) along with two kinds of spontaneous symmetry breaking (blue and red). Gray regions are not connected to the Gaussian UV fixed point. 

In the left panel of Fig.~\ref{PhaseDiagram}, we show the result of this analysis for a representative theory with the two IR fixed point phase structure.  We see that practically all directions connected to the UV, lead to spontaneous symmetry breaking. On the separatrices of the two symmetry breaking regions, we find fine-tuned solutions reaching the IR fixed points, and solutions crossing the intersection of the two symmetry breaking lines.

In the right panel of Fig.~\ref{PhaseDiagram}, the phase diagram for the other phase structure with four IR fixed points is shown. Here we have all three IR phases present.

We notice that the phase diagrams are basically identical to the UV picture in Fig.~\ref{AngularPlot}. From this we conclude, that the lowest order approximations to the symmetry breaking conditions together with the one-loop beta functions are good indicators for the IR fate of CAF theories. This is tightly connected to the relation between fixed flows in the UV and fixed points in the IR. Moreover, the CAF conditions require the couplings to reach fixed flows in the UV, and at this point the symmetry breaking conditions are well approximated by the leading order result. 

For values of the gauge coupling larger than the one at the IR fixed points, other phases exists. However, these are not UV free and are therefore outside the focus of this work.
\begin{figure}[t]
	\centering
	\includegraphics[width=0.4\columnwidth]{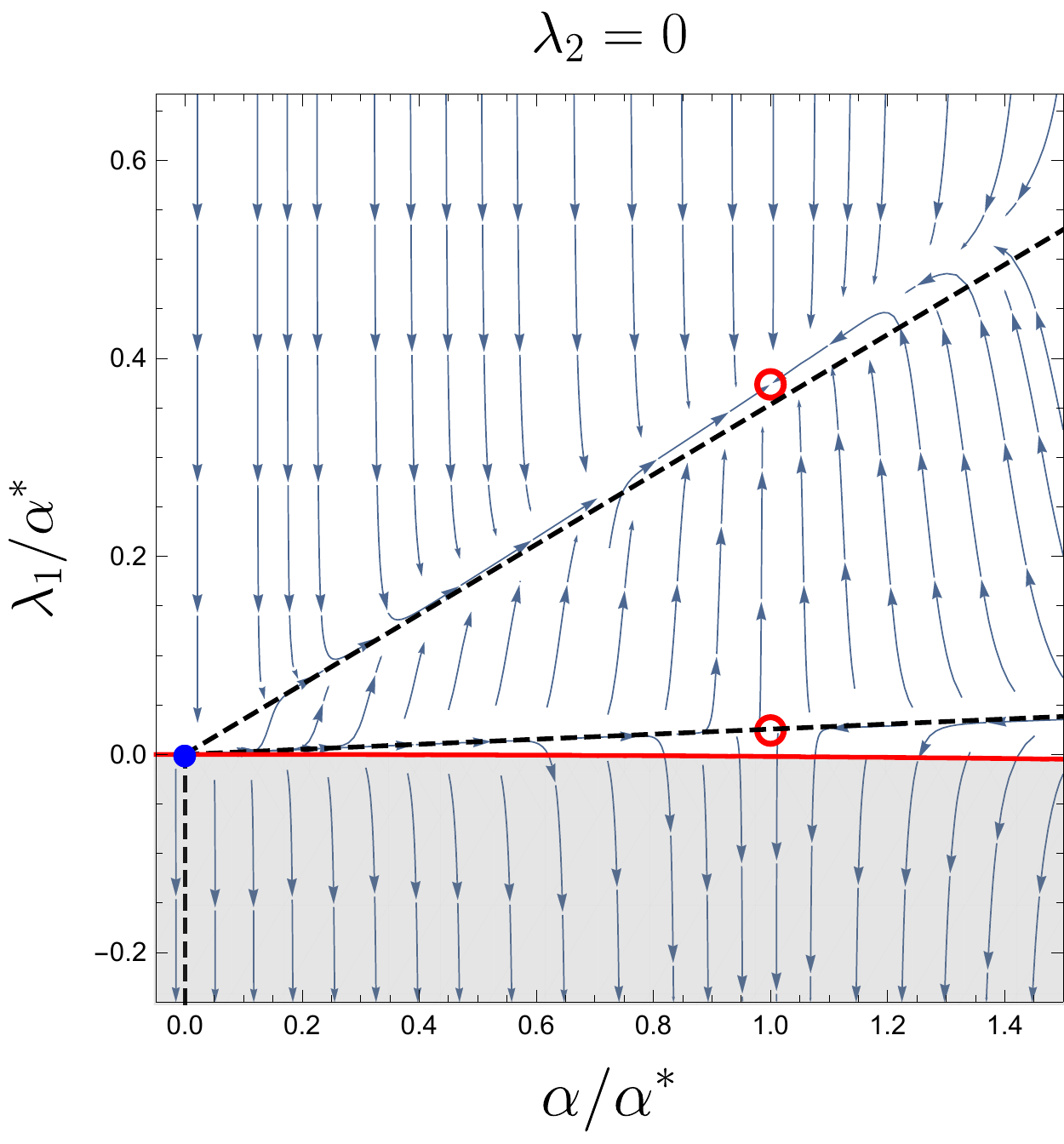}
	\hspace{0.1\columnwidth}
	\includegraphics[width=0.4\columnwidth]{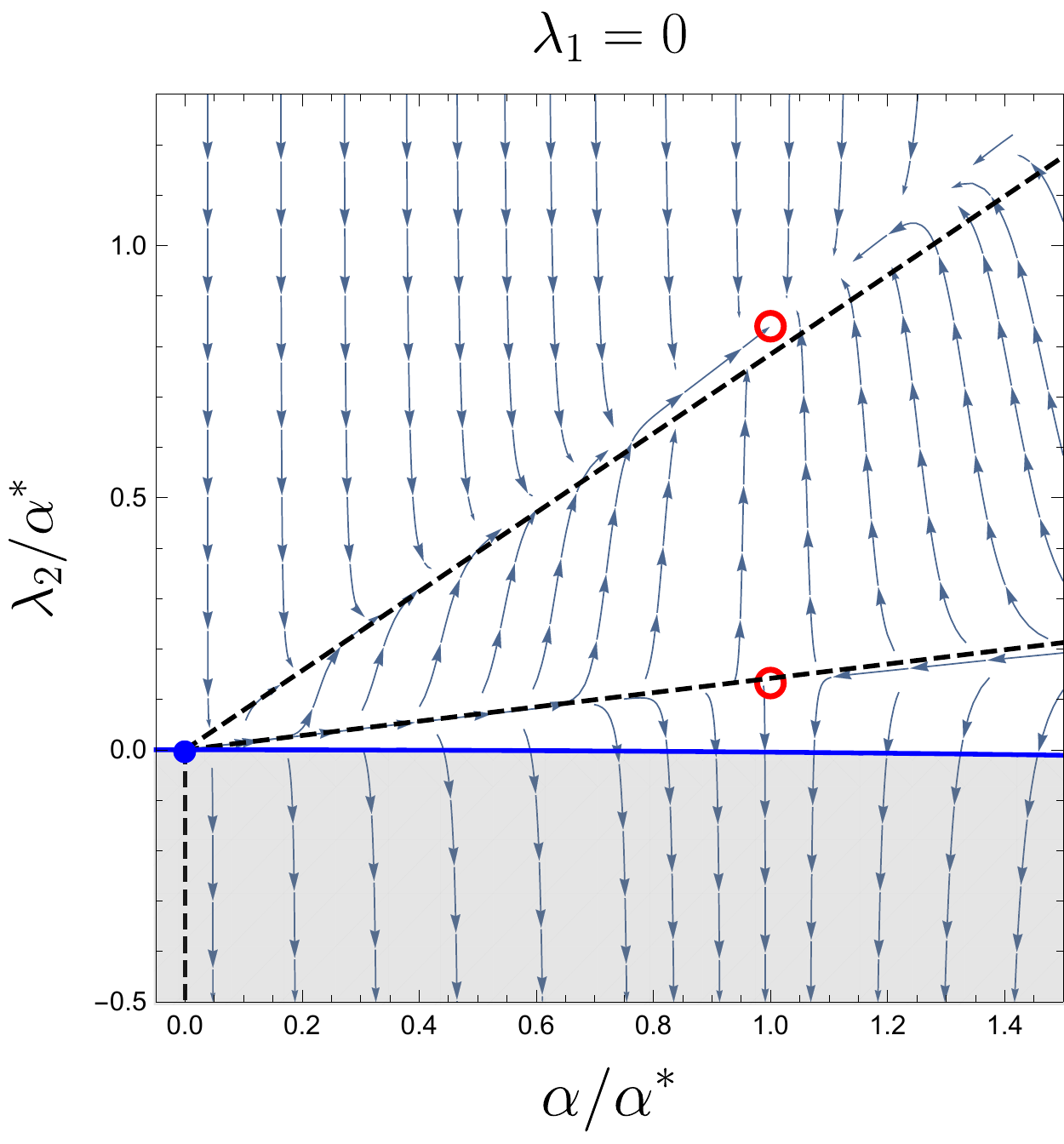}\vspace{0.01\columnwidth}
	\includegraphics[width=0.4\columnwidth]{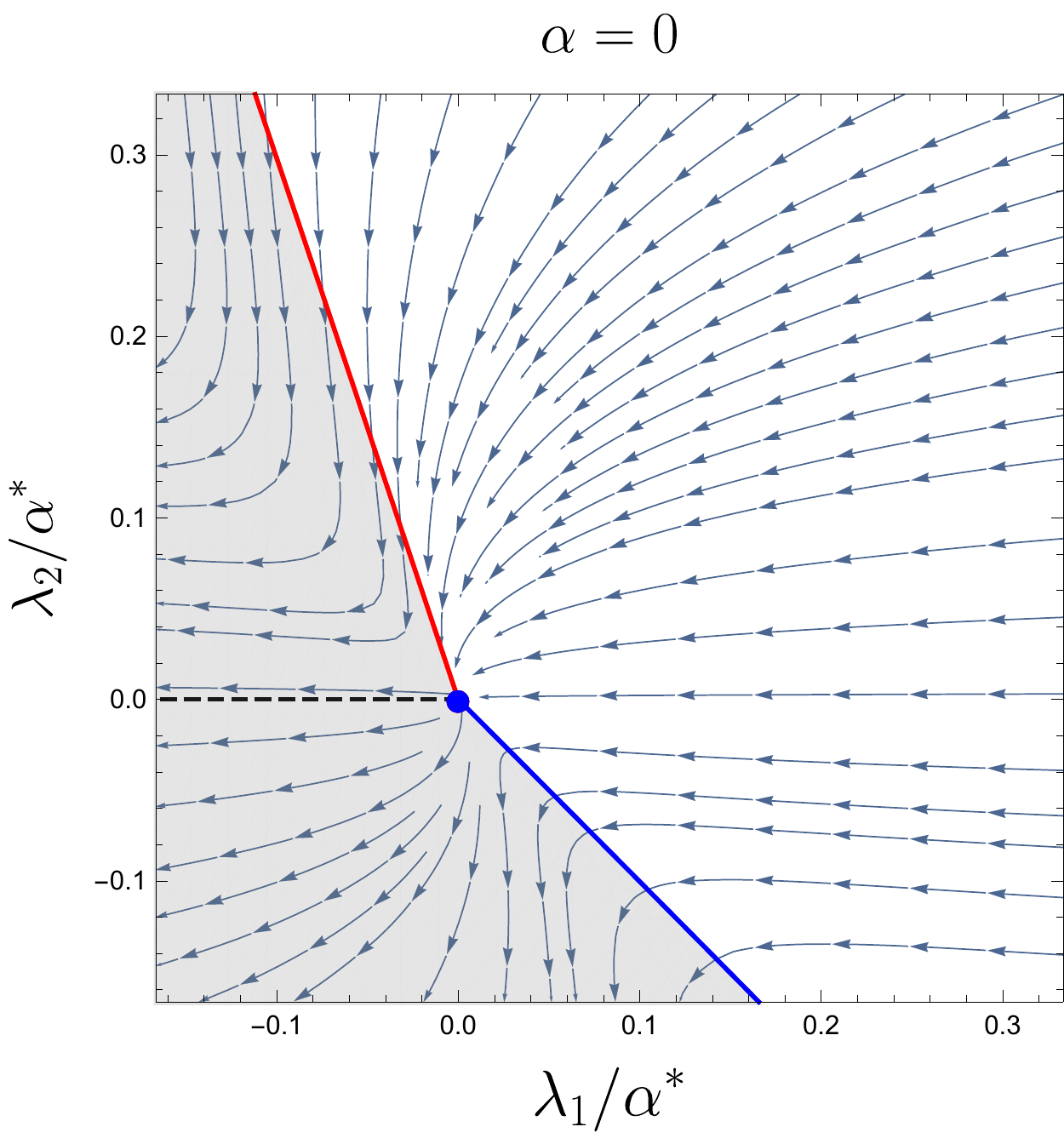}
	\hspace{0.1\columnwidth}
	\includegraphics[width=0.4\columnwidth]{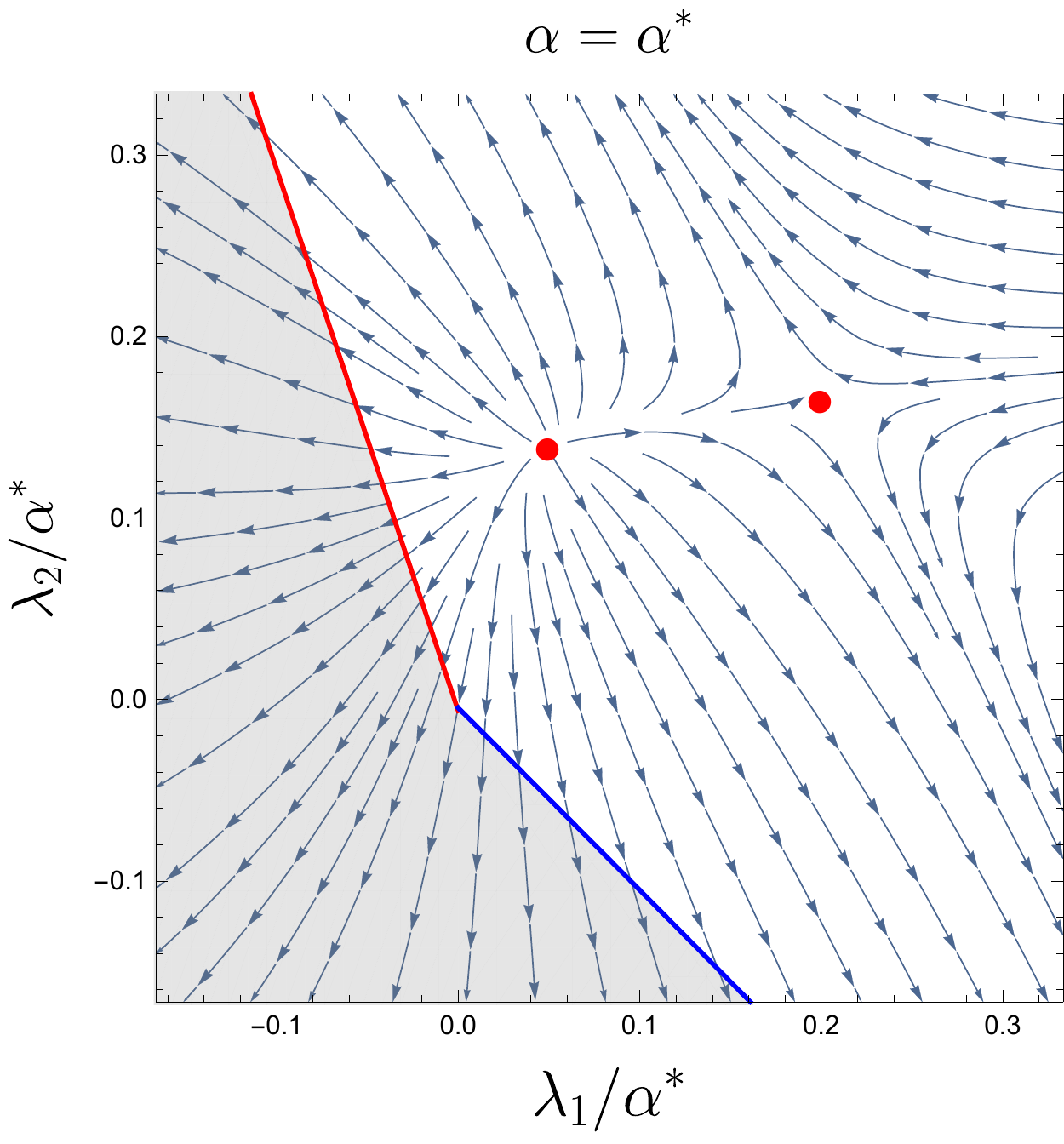}
	\caption{Projected RG flow of the couplings onto the $(\alpha, \lambda_i)$-planes (upper left and right) and $(\lambda_1, \lambda_2)$-plane with the third coupling fixed to $\alpha = 0$ (lower left) and $\alpha=\alpha^*$ (lower right). In each panel we show the lines where spontaneous symmetry breaking occurs (red and blue) including one-loop effects. The gray regions here are the broken phases. Solid dots are fixed points in the full system, while circles mark fixed points in the reduced systems with one coupling kept fixed. Dashed lines mark the fixed flow lines for the reduced systems.}
	\label{PlaneTrajectories}
\end{figure}

In Fig.~\ref{PlaneTrajectories}, we show the RG flow of the couplings projected onto the $(\alpha, \lambda_i)$-planes and $(\lambda_1, \lambda_2)$-plane with the third coupling fixed. In each panel we show the lines where spontaneous symmetry breaking occurs using the one-loop result form Sec.~\ref{loop level eff-analysis}. In the upper left panel, it is clear that the flows above the fixed flow line are not originating from the UV Gaussian fixed point (blue dot), whereas the flows below (unless in the broken phase, marked with gray) all emanate from the fixed point along the other fixed flow line. The same features are seen in the upper right panel. In the lower left panel, we see that only flows in the broken phase originate from the Gaussian UV fixed point, when the gauge coupling is kept fixed, $\alpha = 0$. The lower right panel, shows the two IR fixed points on the plane $\alpha = \alpha ^*$. We see that the symmetry breaking lines both with $\alpha=0$ and $\alpha= \alpha^*$ seem to be identical. This illustrates that for couplings less than or comparable to $\alpha^*$, the tree level result is still a good approximation.  From these diagrams it is clear that the fixed points are all in the unbroken phase.

We have so far uncovered the conditions for the theories to be complete asymptotically free and shown numerically that these conditions are stronger than the conditions for the existence of IR fixed points. Furthermore, we have described the phase structure of the theories anticipating three infrared phases (two types of radiative symmetry breaking and a phase of long distance conformality) using conditions which will be derived in the following section.

\section{Symmetry Breaking}\label{symmetrybreaking}
In order to understand the infrared phases of the theories we now address the question of radiative stability of the scalar potential. In this section we will derive the tree-level stability conditions (flat directions) for the relevant vacuum configurations and derive their corresponding symmetry breaking patterns. Afterwards, we will address the same problem beyond the tree-level analysis,  referring the reader to the appendix for the computational details. 
%
%
%

\subsection{Tree Level Analysis}\label{tree-analysis}
The tree level analysis is the limiting case when the gauge contributions are turned off (i.e.~$\alpha=0$) and higher order terms proportional to $\lambda^n\,\left(n>1\right)$ are ignored. 
Thus, the boundary of the broken phase is a line in the $\lambda_1-\lambda_2$ coupling space rather than a 2 dimensional surface in the $\lambda_1-\lambda_2-\alpha$ space.

The tree level potential has the following form:
\begin{equation}
V=m^2\Tr S^{\dagger}S+\lambda_1\left(\Tr S^{\dagger}S\right)^2+\lambda_2\Tr\left(S^{\dagger}S\right)^2\,,
\end{equation}
where the $N_c\times N_s$ scalar field matrix $S$ is invariant under $SU(N_c)\times U(N_s)$ rotations. 
To illustrate the symmetry breaking pattern, it will be useful to write the matrix $S$ in the form:
  \begin{equation}
    S(x) = U_c^\dagger(x) D(x) U_s(x)\label{diagonalize}
  \end{equation}
  where $U_c$ and $U_s$ are unitary matrices and $D$ is a matrix which is diagonal in a $N_s\times N_s$ block (with real coefficients) and zero everywhere else, assuming that $N_c > N_s$ (which is true for the CAF conditions to be satisfied). Although this is a well known result we summarize the proof in Appendix \ref{diagonalisation}.
    Thus the potential can be rewritten as
  \begin{equation}
   V=m^2\Tr D^{\dagger}D +\lambda_1\left(\Tr D^{\dagger}D\right)^2+\lambda_2\Tr\left(D^{\dagger}D\right)^2\,,
  \end{equation}
  i.e. all the dependence on the unitary matrices $U_c$ and $U_s$ vanishes.
  The 1-loop effective potential
will likewise depend only on the components of $D$ and so only the  components of $D$ can obtain non-zero vacuum expectation values.

 The relevant degrees of freedom relevant in understanding the behaviour of the effective potential are the diagonal part diag$\left(\rho_1,\rho_2,\cdots,\rho_{N_s}\right)$ of $D$. 
The tree level potential can thus be simplified into the following form:
\begin{equation}
V=m^2\sum_{i=1}^{N_s}\rho_i^2+\lambda_1\left(\sum_{i=1}^{N_s}\rho_i^2\right)^2+\lambda_2\sum_{i=1}^{N_s}\rho_i^4\,,\label{potential simplified}
\end{equation}
where $m^2$ could be positive, zero or negative. 
For a positive mass term, i.e.~$m^2>0$, the potential has minimum at $\rho_i = 0$, and this excludes symmetry breaking, while for $m^2<0$, we will have spontaneous symmetry breaking as long as the potential is bounded. One can further show that the non-trivial vacuum configurations for the $m^2 < 0$ case are the same as in the massless case dictated by the sign of $\lambda_2$.  

Restricting to the $m^2=0$ case\footnote{The massless case corresponds to classically conformal models which possess many interesting features such as providing naturally light Higgs in asymptotically safe or free scenarios \cite{Pelaggi:2017wzr}  (see also earlier works \cite{Weinberg:1978ym,Bardeen,Hill:2014mqa}).}, we will now follow Ref.~\cite{Gildener:1975cj} to determine the rays along which the potential vanishes.
Without loss of generality we constrain the $\rho_i$'s on an $N_s$-dimensional hypersphere, i.e. $ \sum_{i=1}^{N_s} \rho_i^2 = l$. The constraint is imposed on the potential through a Lagrange multiplier $L$, leading to: 
\begin{equation}
V = \lambda_1 l^2 + \lambda_2 \sum_{i=1}^{N_s} \rho_i^4 + L \left( \sum_{i=1}^{N_s} \rho_i^2 - l\right)\,.\label{potential_Lagrangian_1}
\end{equation}
The condition to minimize the potential is then given by:
\begin{equation}
\frac{\partial V}{\partial \rho_j} =4 \lambda_2 \rho_j^3 + 2 L \rho_j = 0\,,
\end{equation}
providing the solutions:
\begin{equation}
\rho_j^2 = - \frac{L}{2\lambda_2} \quad \text{ or }\quad \rho_j^2=0\,.
\end{equation}
It is clear that at the extrema of $V$ on the sphere, all non-zero elements of $\rho_i^2$ will be equal. Suppose there are $k$ non-zero elements with value $\rho$, then we find that $\rho^2 = \frac{l}{k}$, and we obtain:
\begin{equation}
\left.V \right|_{\text{ext}} = l^2 \left(\lambda_1 + \frac{\lambda_2}{k} \right)\,.\label{potential extreme}
\end{equation}
When $\lambda_2 > 0$, the potential attains a minimal value at the extremum with $k$ as large as possible, leading to $k = N_s$ 
\begin{equation}
\left. V\right|_{\text{min}} = l^2 \left(\lambda_1 + \frac{\lambda_2}{N_s} \right), \quad \text{for } \lambda_2 > 0\,.\label{min1},
\end{equation}
whereas if $\lambda_2 < 0$, the potential will be minimal for $k = 1$, i.e.
\begin{equation}
\left. V\right|_{\text{min}} = l^2 \left(\lambda_1 + \lambda_2 \right), \quad \text{for } \lambda_2 < 0\,. \label{min2}
\end{equation}

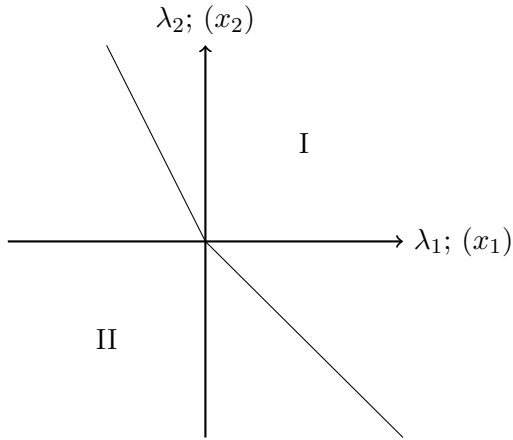
\begin{figure}
\centering
\begin{tikzpicture} [scale = 1.3]
\draw [->, thick] (0,0) -- (0,4) node [above] {$\lambda_2;\,\left(x_2\right)$}; 
\draw [->, thick] (-2,2) -- (2,2) node [right] {$\lambda_1;\,\left(x_1\right)$}; 
\draw (2,0) -- (0,2);
\draw (-1,4) -- (0,2);
\draw (1,3) node {I};  
\draw (-1,1) node {II};   
\end{tikzpicture}
\caption{Boundary lines in the parameter space $\lambda_1 - \lambda_2$ ($x_1 - x_2$), with $x_i = \lambda_i/\alpha$,  across which symmetry breaking can occur. In region I the tree level potential is bounded from below, whereas it is unbounded in region \RNum{2}.}\label{tree level region plot}
\end{figure}

In order for the direction to be flat, we require the $\left. V\right|_{\text{min}} = V(\rho_i = 0) = 0$ to obtain a ray on which loop effects can induce spontaneous symmetry breaking. 
These rays exist under two conditions:
\begin{align}
\text{For } \lambda_2 > 0:&  \qquad N_s \lambda_1+\lambda_2=0\label{broken phase boundary line_1}\\
\text{For } \lambda_2 < 0:&  \qquad \lambda_1+\lambda_2=0\,.\label{broken phase boundary line_2}
\end{align}
These lines are summarized in Fig.~\ref{tree level region plot}.
In the figure we overlap two sets of coordinates $(\lambda_1,\lambda_2)$ and $(x_1 ,x_2)$, with $x_i = \lambda_i/\alpha$, for which the lines coincide. Requiring the potential to be bounded from below, implies the RG flow in the UV ($t\rightarrow\infty$) to be in region \RNum{1}. However for a complete asymptotically free theory, $\lambda_i = 0$. To plot these theories in the diagram, it is useful to use the alternative parameters $x_i$, for which the UV fixed point is not at the origin. This condition was used in Sec.~\ref{CAF} to correctly find the CAF theories. 
For spontaneous symmetry breaking to occur, the RG flow (from UV to IR) must run from region \RNum{1} to region \RNum{2}, irregardless of the chosen set of coordinates. 

We will now discuss the symmetry breaking patterns. These can be directly read from the vacuum configurations as follows.

For $\lambda_2 < 0$, $V$ is minimized when there is only one non-zero $\rho_i$, which leads to the following form of the vacuum configuration:
\[ \langle S_{ia} \rangle = \rho \delta_{i1} \delta_{a1}\,.\]
The corresponding symmetry breaking pattern is
  \begin{equation}
  	SU(N_c)\times U(N_s) \to SU(N_c-1) \times U(N_s-1) \times U(1).\label{SB pattern 1}
  \end{equation}
For $\lambda_2 > 0$, $V$ is minimized when there are $N_s$ non-zero $\rho_i$ providing:
\[ \langle S_{ia} \rangle = \rho \delta_{ia}\,.
\]
In this case the symmetry breaking pattern is
\begin{equation}
	SU(N_c)\times U(N_s) \to SU(N_c-N_s) \times SU(N_s) \times U(1).\label{SB pattern 2}
\end{equation}
These symmetry breaking patterns are worked out in detail in App.~\ref{app:symmetrypatterns}.

This concludes the tree level analysis, naturally leading to the question as to whether or not higher orders can affect these findings. This will be discussed momentarily. 

\subsection{Quantum corrections}\label{loop level eff-analysis}  

 At the quantum level interesting possibilities for the vacuum structure of the theory may emerge when scalar and gauge couplings start competing. However in the deep UV the scalar couplings $\lambda_i$ scale linearly with $\alpha$ and therefore in this region the classical analysis remains intact. This is in line with the expectation that radiative corrections (due to the gauge coupling)  start becoming relevant for $\lambda_i \sim \alpha^2$.  However if the RG running of the couplings crosses the line Eq.~(\ref{broken phase boundary line_1}) or (\ref{broken phase boundary line_2}), while  $\lambda_i \gg \alpha^2$, quantum corrections can induce spontaneous symmetry breaking along the corresponding tree level flat directions derived in Sec.~\ref{tree-analysis}. 
 
As it can be seen from the UV picture in Fig.~\ref{AngularPlot}, there are UV free trajectories which run arbitrarily close to the origin in $(\lambda_1,\lambda_2)$-space (indicated by a black dot), where the inequality $\lambda_i < \alpha^2$ holds. In these cases, the tree-level potential is approximately flat in all directions, and the symmetry breaking is dominated by the gauge loop-contributions.  
%
%
%
The symmetry breaking is therefore no longer restricted to the symmetry breaking patterns discussed in the previous section. This possibility was firstly pointed out in \cite{Gildener:1975cj}. This scenario is phenomenologically interesting, since when all directions are flat, the theory could radiatively generate a spectrum for all the scalar masses (see e.g.~\cite{Hill:2014mqa,Wang:2015cda,Steele:2013fka}), as compared to the Gildener-Weinberg scenario \cite{gildener}, where the only radiatively generated scalar mass is the mass of the scalon.

There are two main ways to implement the one loop effects: explicit logarithmic summation (see e.g.~\cite{Wang:2015cda,Steele:2013fka,Wang,Steele:2014dsa}) or implicit logarithmic summation (see e.g.~\cite{Casas:1994us}). In the latter case, the renormalization group improved effective potential is:
\begin{equation}
  V\left(\{\rho_i\}\right)=\left[\lambda_1\left(t\right)\left(\sum_{i=1}^{N_s}\rho_i^2\right)^2+\lambda_2\left(t\right)\sum_{i=1}^{N_s}\rho_i^4\right]\exp\left(4\int_0^t dt'\gamma\left(t'\right)\right) \,,\label{RG Improved potential simplified_1}
\end{equation}
where $\gamma\left(t\right)$ is the anomalous dimension of the scalar field and $t$ is defined as $t=\log\left[\sum_{i=1}^{N_s}\rho_i^2/\mu^2\right]$ with renormalization scale $\mu$. 
At one loop level, the anomalous dimension depends only on the Yukawa couplings, which are not present in our model, and we can set $\gamma\left(t\right)=0$. We can therefore further simplify the RG improved effective potential, Eq.~\eqref{RG Improved potential simplified_1}, to:
\begin{equation}
  V\left(\{\rho_i\}\right)=\lambda_1\left(t\right)\left(\sum_{i=1}^{N_s}\rho_i^2\right)^2+\lambda_2\left(t\right)\sum_{i=1}^{N_s}\rho_i^4 \equiv \lambda_1(t) f_1({\rho_i})+\lambda_2(t) f_2({\rho_i})\,,\label{RG Improved potential simplified_2}
\end{equation}
 where we introduced functions $f_1$ and $f_2$ for convenience.

The RG improved minimization condition is given by:
\begin{equation}
  V_{\rho_i}^{\left(1\right)}\equiv\frac{\partial V}{\partial\rho_i} = \sum_{j=1}^{2} \left( \frac{d\lambda_j}{dt}\frac{\partial t}{\partial \rho_i} f_j + \lambda_j \frac{\partial f_j}{\partial \rho_i} \right) = \sum_{j=1}^{2} \left( \beta_{\lambda_j} \frac{\partial t}{\partial \rho_i} f_j + \lambda_j \frac{\partial f_j}{\partial \rho_i} \right)\,.\label{RG improved VEV condition}
\end{equation} 
Now it is clear that although the effective potential is of the form of the tree level potential, the one loop information is encoded through the RG functions ($\beta_{\lambda_1},\beta_{\lambda_2}$). 
 
In order for alternative vacuum configurations to exist, we need either a number of distinct values of $\rho_i$ to satisfy Eq.~(\ref{RG improved VEV condition}), or the minimum to be at different number of non-zero $\rho_i$'s than at tree level. For each distinct non-zero value of $\rho_i$, the minimization condition corresponds to a non-trivial constraint on the couplings. Since we have three marginal couplings $\left(\alpha ,\lambda_1,\lambda_2\right)$, we therefore expect that at most there could be two distinct vacuum expectation values to fully determine symmetry breaking lines in $(\alpha, \lambda_1, \lambda_2)$-space. For three distinct vacuum expectation values, all three couplings will be fully determined. An exception to this is when $\lambda_2 = 0$; then there is only a single constraint on the remaining couplings which depends on $f_1$ and $f_2$.

We assume in the following that the vacuum configuration to be such that $n_1$ values of $\rho_i$ are $\rho$, $n_2$ are $\kappa \rho$, with $\kappa$ positive and different from unity, and $N_s - n_1 - n_2$ values equal to zero. From Eq.~(\ref{RG improved VEV condition}) we get two constraints on the couplings. 

Furthermore, the effective potential needs to be stable at the vacuum configuration, we therefore derive the eigenvalues of the Hessian matrix
\begin{equation}
\left. V_{\rho_i\rho_j}^{\left(2\right)}\equiv\frac{\partial^2V}{\partial\rho_i\partial\rho_j}\right|_\text{vacuum}\, ,\label{Hessian matrix}
\end{equation}  
where all three RG functions ($\beta_{\lambda_1},\beta_{\lambda_2},\beta_\alpha$) are encoded. 
A stable vacuum and physical scalar masses requires the  eigenvalues to be non-negative. 

We find that these requirements cannot be met unless $n_2 = 0$, and $n_1$ is either 1 or $N_s$, which are exactly the tree level vacuum configurations, implying no alternative symmetry breaking patterns are found in this model beyond the two discussed at the tree level analysis, i.e. Eq.~\eqref{SB pattern 1} and Eq.~\eqref{SB pattern 2}. 

In App.~\ref{app:RGimproved}, we carry out the analysis of the case with $\left(N_c=6, N_s=3, N_f=31\right)$.
We find that the RG improved boundary lines \eqref{RG improved boundary 1} and \eqref{RG improved boundary 2}  for the broken phases actually shift the tree level boundary lines when $\lambda_i \sim \alpha^2 \ll 1$. While, when $\lambda_i \sim \alpha \ll 1$ the RG improved reduces to the tree level result. In this way, Fig.~\ref{Region Plot_3} provides a detailed view of region near the origin of Fig.~\ref{tree level region plot}.

Similarly, in App.~\ref{eff-explicit}, we perform the same analysis based on the explicitly calculated one-loop effective potential in  the Coleman-Weinberg renormalization scheme. There are differences in the exact conditions  on the couplings for spontaneous symmetry breaking, but the corresponding vacuum configurations, and thus symmetry breaking patterns, are identical and the regions satisfying the vacuum stability conditions (comparing Fig.~\ref{Region Plot_X} with Fig.~\ref{Region Plot_1}) are similar and consistent in both renormalization schemes.

\section{Summary}
We have simultaneously carried out a detailed study of the UV behaviour and a classification of the IR phases of $SU(N_c)$ gauge theories with $N_s$ complex scalars and $N_f$ vector-like fermions in the fundamental representation.
 
This entailed a careful analysis of the conditions for complete asymptotic freedom (CAF). Interestingly, due to the presence of 
  fundamental scalars,  CAF requires a large number of colors, $N_c$, and a large number of fermions. We find for specific combinations of $N_c$ and $N_s$ a window in $N_f$ for which the CAF conditions are satisfied. The most minimal case is $N_c = 5$ with $N_s = 2$. The CAF allowed number of fermions,  $N_f$, is found to be close to the loss of asymptotic freedom in the gauge beta function. We show that the CAF conditions are, remarkably,  more restrictive than the requirement for the theory to have IR fixed points, when considering higher orders. This means that any CAF theory of this kind displays long distance conformality, at least in some coupling direction. We stress that our results are within perturbative control.
 
When considering the infrared fate of the theory, we discover two distinct phase structures. For most combinations of $N_c$, $N_f$ and $N_s$ we have two IR fixed points, while for larger values of $N_c$ and $N_f$ with $N_s$ small,  four IR fixed points exist. 
For the theories featuring two IR fixed points, neither of them are fully IR attractive and furthermore they reside on the separatrix between two radiatively broken phases. 
However for theories featuring the four IR fixed points we observe that a fully IR attractive fixed point appears allowing for a stable   phase of long distance conformality.

To investigate the possible existence of radiative symmetry breaking, we performed analyses both at tree- and one- loop levels.
For the tree level analysis, we used the conventional Gildener-Weinberg method, while at the quantum level, we used the renormalization group improved effective potential.
 The same two symmetry breaking patterns were found for both analyses: $SU(N_c) \times U(N_s) \rightarrow SU(N_s) \times SU(N_c - N_s)\times U(1)$ and $SU(N_c) \times U(N_s) \rightarrow SU(N_c-1) \times U(N_s -1) \times U(1)$. This is despite the fact that the loop level analysis allows one to study regions of phase space, where quantum corrections are dominating the vacuum configuration of the scalar fields. 

Our analysis has shed light on the UV behaviour and rich low energy phase structure of minimal extensions of QCD-like theories featuring scalar quarks. We discovered 
that ensuring  these theories   to be fully asymptotically free is related to the presence of long-distance conformality.  Our results can be useful when constructing extensions of the standard model featuring  vector-like dynamics.

\acknowledgments

This work is partially supported by the Danish National Research Foundation under the grant DNRF:90. 
T.G.S. and R.B.M are grateful for financial support from the Natural Sciences and Engineering Research Council of Canada (NSERC). Z.W.~Wang thanks Matin Mojaza and Esben Molgaard for helpful discussions.

\newpage
\appendix

\section{Quartic polynomial and classification of its roots}\label{quartic}

Here we provide the details for the procedure outlined in   Sec.~\ref{CAF} of reducing the fixed flow equation (\ref{eq:fixedflow}) to a quartic equation in a single coupling, $\lambda_2$. We will use the discriminant method to classify the roots of the equation. 

We want to find solutions to
\begin{equation}
(\beta_{\alpha} , \beta_1, \beta_2) = c ( \alpha, \lambda_1, \lambda_2)\, ,
\label{eq:fixedflow2}
\end{equation}
in the case where $c<0$ and the beta functions are given by Eq.~(\ref{betafunctions}). First we note that $\beta_\alpha$ is only a function of $\alpha$ itself. We can therefore find the fixed flow solution for $\alpha\neq 0$ must satisfy $c = - B\alpha$, where $B = \tfrac13 \left(22N_c-4N_f-N_s\right)$. Clearly, the condition $c<0$ is only satisfied when $B>0$.  We can now substitute $c = - B\alpha$ into the remaining components of  Eq.~(\ref{eq:fixedflow2}). For $\alpha\neq 0$ it is convenient to introduce the rescaled couplings, $\lambda_i = \lambda_{is} \, \alpha$, for which we can factor out the gauge coupling dependence of the last two components of Eq.~(\ref{eq:fixedflow2}). In other words, the two equations, $\alpha^{-2}\left(\beta_i - c \lambda_i \right)=0$, can be written as 
\begin{align}\label{eq:reducedflow}
4 (N_c N_s+4) \lambda_{1s} ^2 +12\lambda_{2s}^2+\lambda_{1s} \left[ 8(N_c+N_s)\lambda_{2s} + B - \frac{6(N_c^2-1)}{N_c} \right]+ \frac{3 (N_c^2 + 2)}{4N_c^2} & =0 \, ,\nonumber \\
4(N_c+N_s)\lambda_{2s}^2+\lambda_{2s} \left[24\lambda_{1s} + B  - \frac{6(N_c^2-1)}{N_c}\right]+\frac{3 (N_c^2 - 4)}{4 N_c} & =0\, .
\end{align}
In the case, where $B=0$, the equations above no longer describe solutions to Eq.~(\ref{eq:fixedflow2}), instead they correspond to the equations $\beta_i =0$ for constant $\alpha\neq 0$. These solutions we refer to as pseudo fixed points, since in general the gauge coupling is not fixed.

Now, for $B>0$, we can solve the second equation in Eq.~(\ref{eq:reducedflow}) for $\lambda_{1s}$ and substitute into the first in order to obtain a quartic polynomial in $\lambda_{2s}$ (after multiplying with $\lambda_{2s}^2$)
\begin{equation}\label{eq:quartic}
a\, \lambda_{2s}^4 + b\, \lambda_{2s}^3 + c\, \lambda_{2s}^2 + d\, \lambda_{2s} + e = 0\, ,
\end{equation}
where 
\begin{align}
9 a  &= N_s \left(N_s \left(2 N_c^2+N_c N_s-8\right)+(N_c-4) N_c (N_c+4)\right)-8 N_c^2+108 \, ,\nonumber \\
{18 N_c} b &= (N_c (B-6 N_c)+6) \left(N_s+N_c\right) \left(N_c N_s-5\right) \, ,\nonumber \\
{144 N_c^2} c &= 144 -2 N_c \left(B^2 N_c-12 B \left(N_c^2-1\right)+6 N_c \left(7 N_c^2-25\right)\right) \nonumber \\
&\phantom{{}={}}+N_c N_s \left(\left(B^2-108\right) N_c^2-12 B N_c^3+12 B N_c+42 N_c^4+6 \left(N_c^2-4\right) N_c N_s+84\right) \, ,\nonumber \\
{96 N_c^2} d &= \left(N_c^2-4\right) (N_c (B-6 N_c)+6) \left(N_c N_s+1\right) \, ,\nonumber \\
{256 N_c^2} e &= \left(N_c^2-4\right)^2 \left(N_c N_s+4\right) \, .
\end{align}
In Sec.~\ref{CAF}, we define the critical number of fermions, $N_f^*$, for which $B=0$. Then defining $N_x=N_f^*-N_f$, we can express  $B = 4 N_x / 3$. Following Ref.~\cite{mathbook1}, the nature of the roots of a quartic equation of the form Eq.~(\ref{eq:quartic}) is described by the following functions
\begin{align}
\Delta &= 256 a^3 e^3 - 192 a^2 b d e^2 - 128 a^2 c^2 e^2 + 144 a^2 c d^2 e - 27 a^2 d^4 \nonumber \\
&\phantom{{}={}}+ 144 a b^2 c e^2 - 6 a b^2 d^2 e - 80 a b c^2 d e + 18 a b c d^3 + 16 a c^4 e \nonumber \\
&\phantom{{}={}}- 4 a c^3 d^2 - 27 b^4 e^2 + 18 b^3 c d e - 4 b^3 d^3 - 4 b^2 c^3 e + b^2 c^2 d^2 \nonumber \\
P &=  8ac - 3b^2 \nonumber \\
Q &= b^3+8da^2-4abc \nonumber \\
\Delta_0 &= c^2 - 3bd + 12ae \nonumber \\
D &= 64 a^3 e - 16 a^2 c^2 + 16 a b^2 c - 16 a^2 bd - 3 b^4
\end{align}

Since we are interested in real roots, we have the following relevant cases:
\begin{enumerate}[i)]
	\item If $\Delta < 0$, then equation has two distinct real roots. 
	\item If $\Delta < 0$, while $P<0$ and $D<0$ then all four roots are real and distinct.
	\item If $\Delta = 0$, then the equation has a multiple root and several scenarios exist. 
	
	Only when $D=0$, $P >0$ and $Q = 0$ are none of the roots real. 
\end{enumerate}
In Fig.~\ref{Region2}, we show the regions in $N_c$ and $N_s$ for $N_x \in \{0,1,2,3,4\}$. The upper region satisfies condition (i), the lower region satisfies condition (ii), while the borders satisfy the condition $\Delta = 0$. In the range $N_c\in \left[3,20\right]$, $N_s \in \left[2,20\right]$ and $N_x \in \left[0,20\right]$, there are no integer solutions to the last condition.

\section{Connection between fixed flows and fixed points}\label{app:UVIR}

In Section~\ref{UV-IR}, we found that the set of theories that are CAF is a subset of the theories with interacting IR fixed points. This result was based on a numerical study of the two-loop gauge beta function together with the one-loop beta functions for the quartic couplings. The same study with three-loop gauge beta function is done in App.~\ref{app:three-loop}. However restricting to the two-loop gauge beta function, we know that the running of the gauge coupling is independent of the quartic couplings. Within this approximation and inspecting the  gauge coupling in isolation, the statement above appears easy to disprove. Writing the gauge beta function as $\beta_{\alpha}=-B \alpha^2 + C \alpha^3$, the set of theories with AF is characterized by satisfying the condition, $B>0$, while the interacting IR fixed point is realized only when both $B>0$ and $C>0$. In other words, requiring the theory to have an interacting IR fixed point is a stronger condition than for the gauge coupling to be AF. This is the well known result from Caswell, Banks and Zaks \cite{Caswell:1974gg, Banks:1981nn}.

Adding on top of this the beta functions for the quartic couplings, we know from App.~\ref{quartic} that the CAF condition reduces to $B>0$ and at least a real solution to Eq.~(\ref{eq:reducedflow}), while the existence of an interacting IR fixed point requires at least a real solutions to Eq.~(\ref{eq:reducedflow}) with $B=0$ (not a constraint on $N_c$, $N_s$ and $N_f$) with at least one IR attractive direction. The direction towards the Gaussian fixed point will be IR attractive if $B>0$ and $C>0$. Clearly, our numerical findings in Section~\ref{UV-IR} can be restated as follows: the set of $\{N_c,N_s\}$ for which there are solutions to Eq.~(\ref{eq:reducedflow}) with $B>0$ (constraint on $N_f$) is a subset of the set of $\{N_c,N_s\}$ for which there are solutions to Eq.~(\ref{eq:reducedflow}) with $B=0$, and simultaneously the set $\{N_c,N_s, N_f\}$ satisfying the CAF conditions is a subset of the  $\{N_c,N_s, N_f\}$ satisfying $C>0$, such that the seemingly additional constraint is always implied.  

The first inclusion was shown pictorially in Fig.~\ref{Region2} for specific choices of $N_f$ close to $N_f^*$. In the following we will discuss this relation further and investigate the last inclusion in more depth.  

Clearly, $B>0$ is a common condition for both sets, while the CAF condition takes the form Eq.~(\ref{eq:reducedflow}), the existence of fixed points in $\lambda_1$ and $\lambda_2$ for $\alpha \neq 0$, are the solutions to $\alpha^{-2}\beta_i =0$, which can be written as 
\begin{align}\label{eq:pseudo}
4 (N_c N_s+4) \lambda_{1s} ^2 +12\lambda_{2s}^2+\lambda_{1s} \left[ 8(N_c+N_s)\lambda_{2s} - \frac{6(N_c^2-1)}{N} \right]+ \frac{3 (N_c^2 + 2)}{4N_c^2} & =0 \, ,\nonumber \\
4(N_c+N_s)\lambda_{2s}^2+\lambda_{2s} \left[24\lambda_{1s}  - \frac{6(N_c^2-1)}{N_c}\right]+\frac{3 (N_c^2 - 4)}{4 N_c} & =0\, ,
\end{align}
in rescaled couplings. Notice, that this equation is equivalent to Eq.~(\ref{eq:reducedflow}) without the $B$-terms. However, this derives from the fact that fixed points are solutions to $\beta_i =0$, while fixed flows are solutions to $\beta_i = -B\alpha \lambda_i$. In other words, $B=0$ is not a constraint for the fixed point equation, but a limit in which Eq.~(\ref{eq:reducedflow}) reduces to Eq.~(\ref{eq:pseudo}. 

It is not an easy task to show that the set of $\{N_c,N_s\}$ with real solutions to Eq.~(\ref{eq:pseudo}) is bigger than for Eq.~(\ref{eq:reducedflow}) with $B>0$. However, we can make some simple observations.  Consider only  solutions where both $\lambda_1$ and $\lambda_2$ are positive. Then since both equations, evaluated at $\lambda_{i s}=0$, are positive for $N_c>2$, it is clear that there have to be a solution to  $\beta_i =0$ before there can be a solution to $\beta_i = -B\alpha \lambda_i$, since the latter are negative. For the cases where one or both of the two $\lambda_{i s}$'s are negative, this reasoning does not hold. By demanding a stable scalar potential, the case where both $\lambda_1$ and $\lambda_2$ are negative can be discarded. In this way we are left with the cases with one of the quartic couplings being negative. Here we do not need to cross $\beta_i =0$ for both beta functions  to satisfy $\beta_i = -B\alpha \lambda_i$ and for general coefficients of Eq.~(\ref{eq:reducedflow}), this can be realized. However, for the particular $N_c$ and $N_s$ dependence of the coefficients for this model, it is not the case.    
\begin{figure}[t]
	\centering
	\includegraphics[width=0.55\columnwidth]{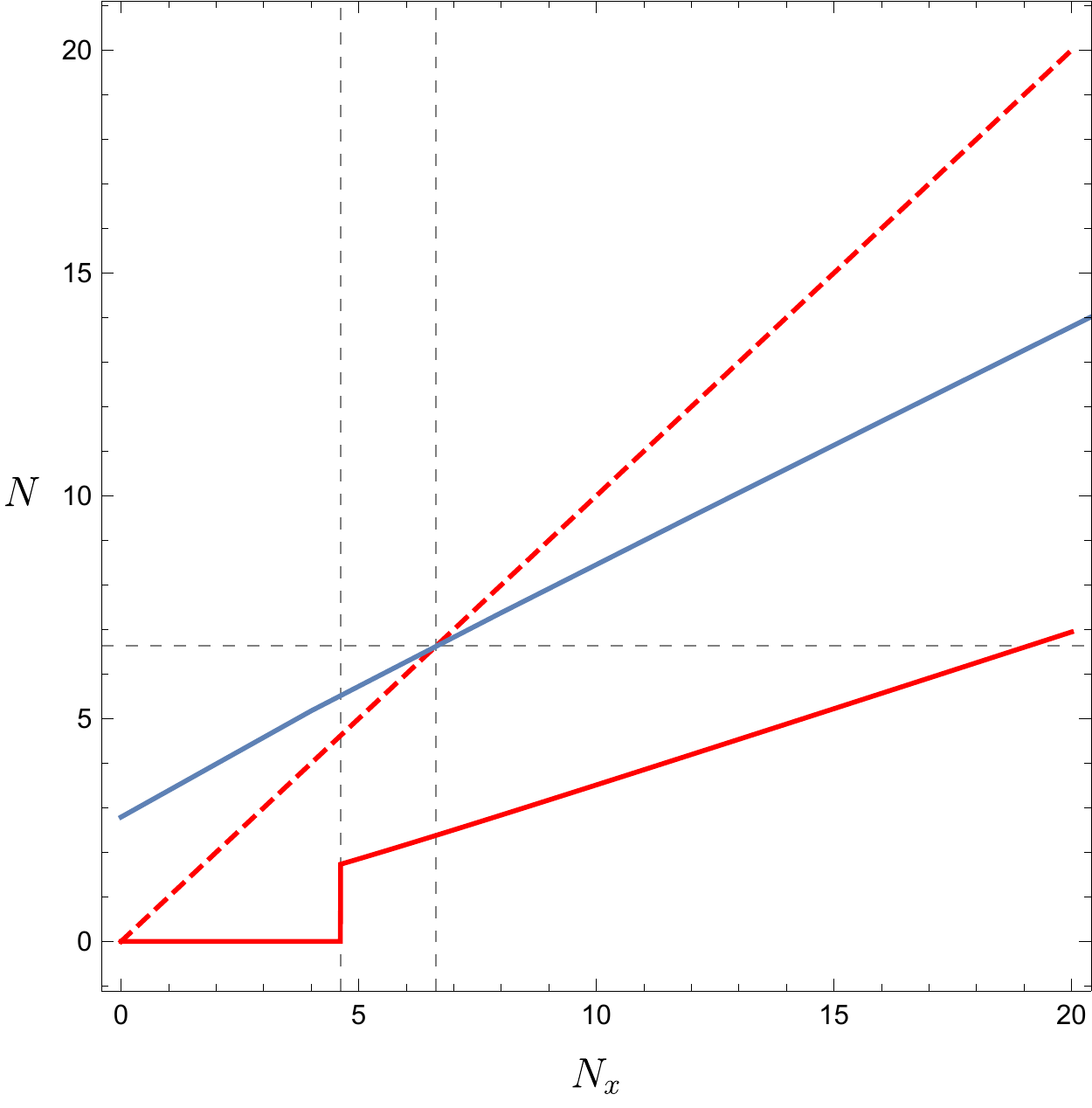}
	\caption{The blue line marks the lower value of $N_c$, as a function of $N_x$, above which there are solutions to Eq.~(\ref{eq:reducedflow}). The red solid line is the value of $N_c$ for which the curve, $C=0$, intersects $N_s = 1$. Below this line there is a region with $N_s >= 1$ for which $C<0$. For $N_x < 8/\sqrt{3}$ the curve with $C=0$ does not intersect $N_s = 1$ (marked with gray dashed line). The dashed red line is straight line $N_c = N_x$ below which the suppression of higher loop contributions to the IR fixed points are of order one.}
	\label{fig:NNx}
\end{figure}
For $0\leq N_f^*- N_f < 8/\sqrt{3}$, the coefficient $C$ is always positive. For $ N_f^*- N_f \geq 8/\sqrt{3}$, the sign of $C$ depends on the $N_c$ and $N_s$. However, since $C>0$ for large values of $N_c$, even when $N_s \sim N_c$, and $C$ is a continuous function in $N_c$, $N_s$ and $N_f$, we can solve for $N_f$ in $C=0$, and check for real solutions to Eq.~(\ref{eq:reducedflow}). Using the same method described in App.~\ref{quartic}, we find no real solutions with $C=0$;  we conclude, since we know that the region with real solutions is connected, that $C>0$ for all obtained solutions. In Fig.~\ref{fig:NNx} we illustrate this fact by plotting the lower value of $N_c$ (blue line), as a function of $N_x$, above which there are solutions, together with the value of $N_c$ (red line) below which $C<0$ for $N_s = 1$. The two lines never intersect, and this supports the statement, that $C>0$ in the whole region of solutions to Eq.~(\ref{eq:reducedflow}).

\begin{figure}[t]
	\centering
	\includegraphics[width=0.4\columnwidth]{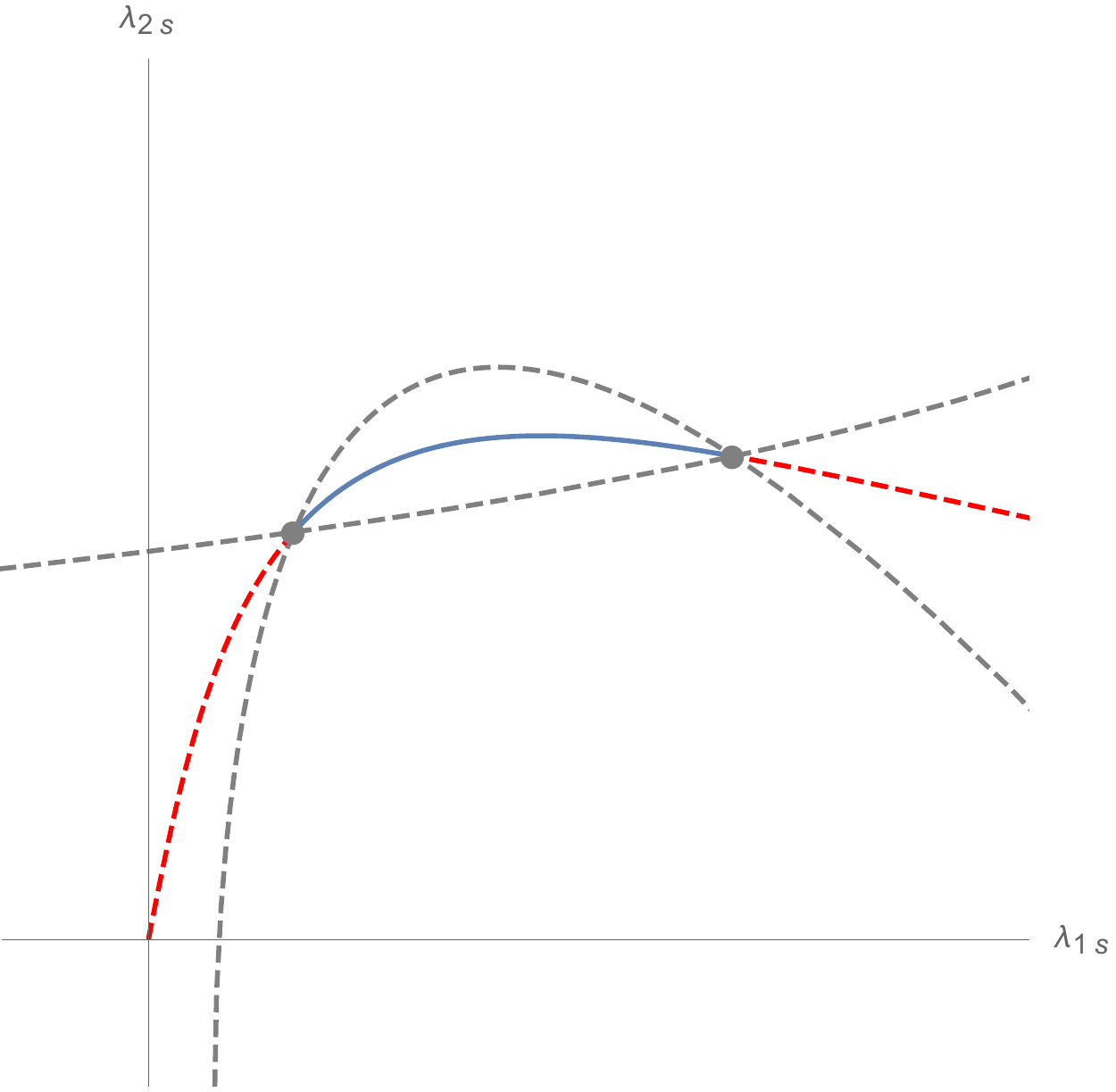}
	\hspace{0.1\columnwidth}
	\includegraphics[width=0.4\columnwidth]{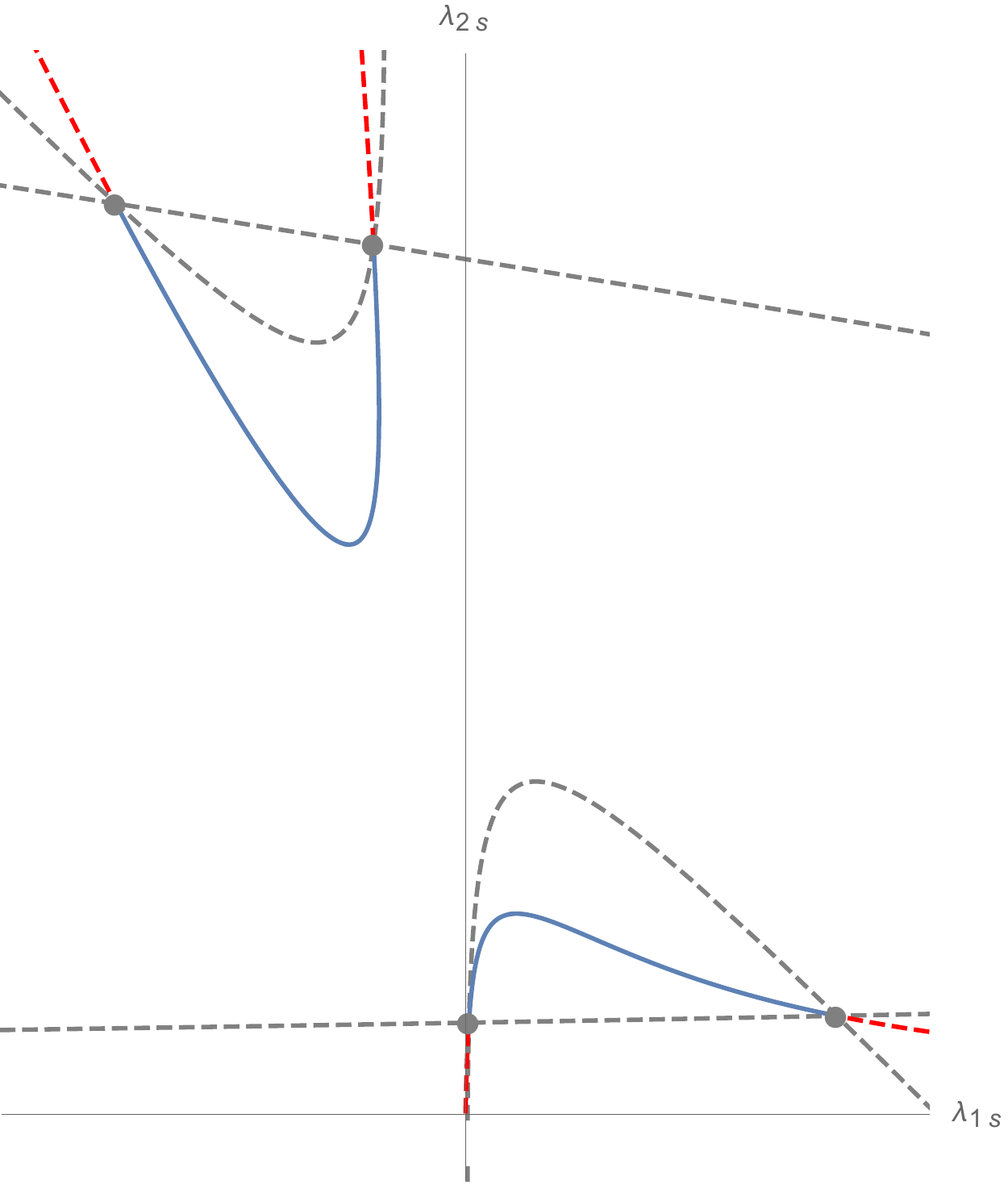}
	\caption{Morphology of the quartic phase diagram. Each gray dashed curve represent the zero-contour of one of the rescaled quartic beta functions. The blue solid lines (red dashed lines) mark the curves where the flow is pointing towards (away from) the origin. The grey dots are the pseudo fixed points. Right: For the case of two solutions. Left: For the case of four solutions.}
	\label{SchematicGeoFlow}
\end{figure}

We will now present the morphology of the phase diagram of the rescaled beta functions given by the left sides of Eq.~(\ref{eq:pseudo}). A study of the curves, where each one of the two beta functions is zero, leads to the conclusion, that in the region where the two distinct real roots exist, both roots will be positive and the phase diagram looks schematically like shown (dashed gray) in left panel Fig.~\ref{SchematicGeoFlow}. In the dark gray region of Fig.~\ref{Region2}, two roots are similar to the previous case and still positive, while the two additional roots in $\lambda_{2s}$ are positive, but larger,, and paired with negative values of $\lambda_{1s}$. The corresponding phase diagram is shown (dashed gray) in the right panel of Fig.~\ref{SchematicGeoFlow}.\\
%
\begin{figure}[t]
	\centering
	\includegraphics[width=0.4\columnwidth]{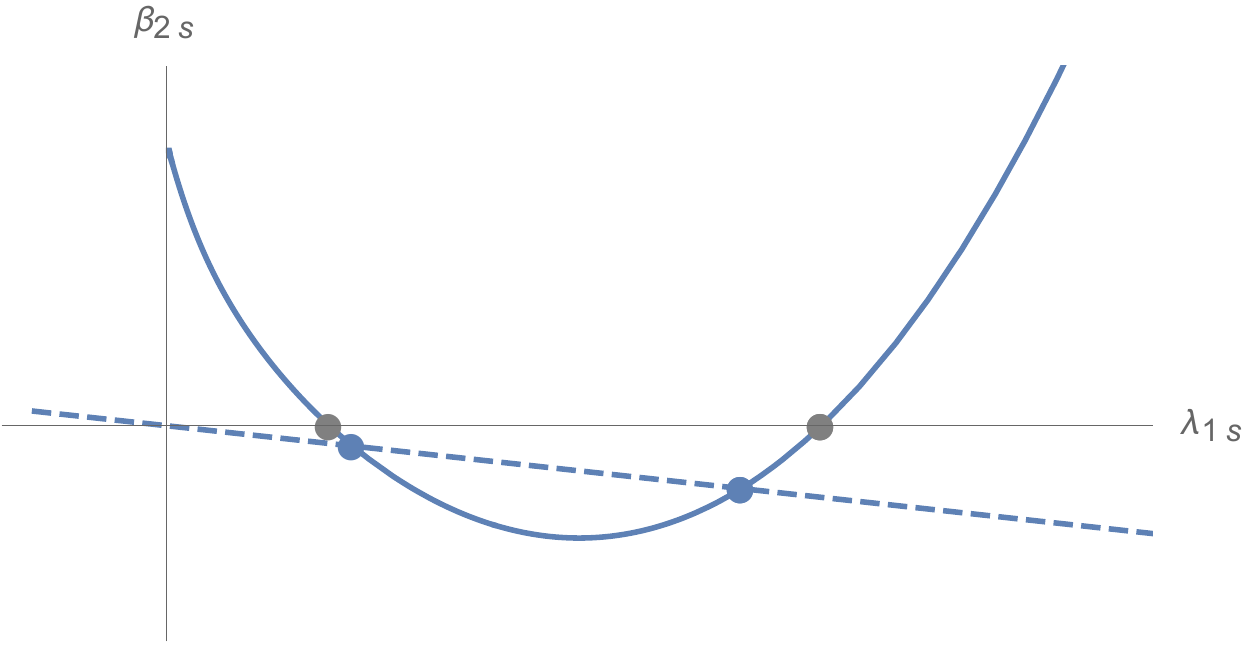}
	\hspace{0.1\columnwidth}
	\includegraphics[width=0.4\columnwidth]{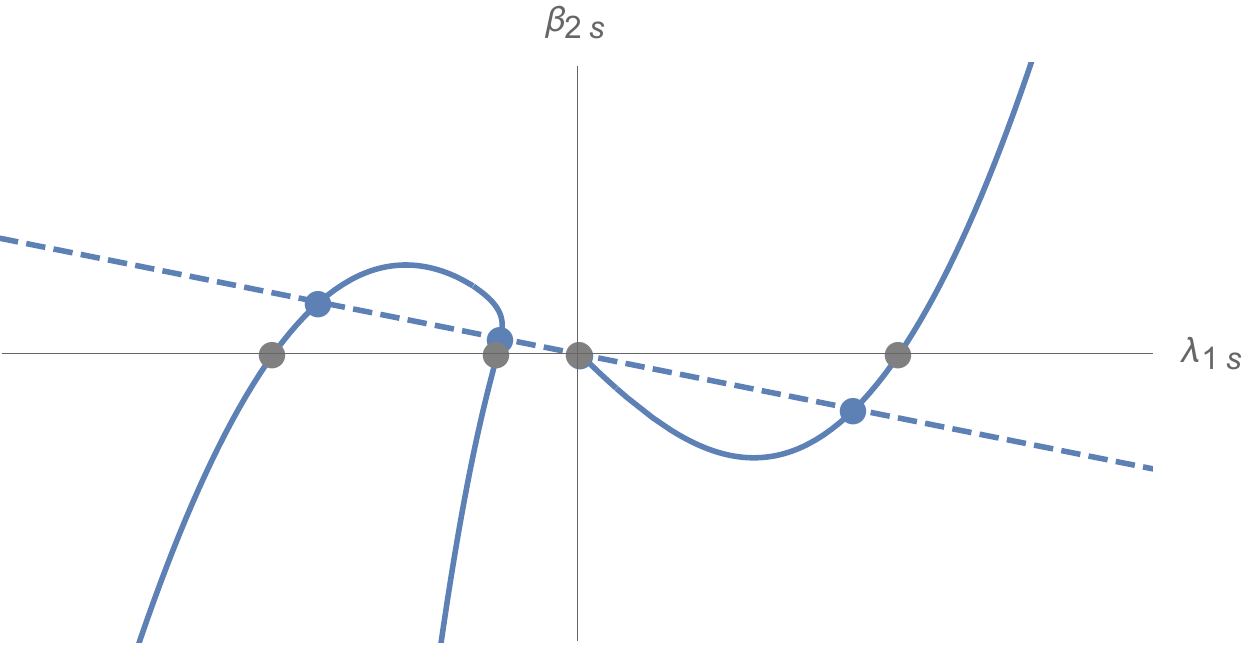}
	\caption{This shows the corresponding beta function of the flows along the lines shown in Fig.~\ref{SchematicGeoFlow}, parametrized by $\lambda_{1s}$. The dashed line is where the $c_s = 2 B$. The grey dots are the pseudo fixed points, and the blue dots are solutions to the fixed flow. Right: For the case of two solutions. Left: For the case of four solutions.}
	\label{SchematicGeoBeta}
\end{figure}
The fixed points (gray dots) are solutions to Eq.~(\ref{eq:pseudo}) where the gauge dependence is factored out. The relation to the actual couplings is $\lambda_i = \lambda_{is} \, \alpha$, and the position of these points will therefore move unless the gauge coupling has a non-trivial fixed point. For theories with CAF, we already know that there exists interacting IR fixed points, $\alpha^*=B/C$.  

In the UV, the gauge coupling is AF and the solutions in $\lambda_{is}$  are pseudo fixed points which all go to zero. If we require the quartic couplings also to be AF, they need to be fixed in this geometrical rescaling of the pseudo fixed points. Possible candidates are the points where the flow is pointing towards the origin, i.e. $(\beta_{\lambda_1},\beta_{\lambda_2}) = c (\lambda_1, \lambda_2)$ with $c<0$, i.e. the last two equations in Eq.~(\ref{eq:fixedflow}). Factoring out the gauge coupling, the condition becomes,
\beq\label{Quartic2}
\alpha^{-2}(\beta_{1},\beta_{2}) = c_s (\lambda_{1s}, \lambda_{2s})\, 
\eeq
where $c_s= c /\alpha$. When $c_s = -B$ this equation equals Eq.~(\ref{eq:reducedflow}). Solving the equation for $c_s<0$ gives the blue solid lines in  Fig.~\ref{SchematicGeoFlow}, whereas the solutions for $c_s>0$ are shown as dashed red lines.

In order to leave the geometry with respect to the pseudo fixed points unchanged, the magnitude of the flow needs to match the change in the gauge coupling, $\alpha$, i.e. corresponding to a uniform contraction. 
\beq
(\beta_\alpha, \beta_{\lambda_1},\beta_{\lambda_2}) =  c (\alpha, \lambda_1, \lambda_2)
\eeq
This condition is the fixed flow equations Eq.~(\ref{eq:fixedflow}) for the gauge and quartic couplings introduced in Sec.~\ref{CAF}. Here we showed that $c = - B\, \alpha$. This means that the solutions to the CAF condition are specific points along the blue curves. In Fig.~\ref{SchematicGeoBeta}, we plot the rescaled beta function, $\beta_{2 s} = \alpha^{-2} \beta_{2}$, for $\lambda_2$ along the line given by Eq.~(\ref{Quartic2}) parametrized by $\lambda_{1s}$. In the same picture, we superimpose the condition $c \lambda_2$ for $c<0$. From this plot, we see the relation between the fixed flow solutions (blue dots) and the pseudo fixed points (gray dots). Furthermore, we see why they come in pairs, and since we assume the shape to be characteristic for the whole solution space, we understand why we always have pseudo fixed points when we have fixed flow solutions.

\section{Three loop gauge contribution analysis}\label{app:three-loop}

As discussed in section~\ref{IRbehaviour}, for a Weyl-consistent approach we would have to include the three loop correction to the gauge renormalization group equation in our IR analysis. This takes the form:
\begin{align}\label{betag3}
\beta_{\alpha}^{(3)} &= (-2N_s \lambda_1^2 - 2N_c N_s^2 \lambda_1^2 - 4 N_c N_s \lambda_1 \lambda_2 - 4 N_s^2 \lambda_1 \lambda_2 - 2N_s \lambda_2^2 - 2N_c N_s^2 \lambda_2^2) \alpha^2\\ \nonumber
&\quad + \left( N_c N_s \lambda_1 - \frac{2N_s \lambda_1}{N_c} + 2N_s^2 \lambda_1 + 2N_s \lambda_2 - \frac{2 N_s^2 \lambda_2}{N_c} + N_c N_s^2 \lambda_2 \right) \alpha^3\\ \nonumber
& \quad + \Big( \frac{1709 N_c^2 N_f}{27}-\frac{2857 N_c^3}{27} - \frac{187 N_f}{18} - \frac{N_f}{2N_c^2}  + \frac{11 N_f^2}{9 N_c} - \frac{102 N_c N_f^2}{27}- \frac{1651 N_s}{77}\\ \nonumber
& \quad \qquad + \frac{29N_s}{8N_c^2}+\frac{1315 N_c^2 N_s}{56} + \frac{73 N_f N_s}{36 N_c} - \frac{335 N_c N_f N_s}{108} + \frac{49 N_s^2}{77N_c} - \frac{143 N_c N_s^2}{216} \Big) \alpha^4
\end{align}

We notice that the gauge coupling is no longer decoupled from the quartic coupling system, which makes a similar approach to the one performed in section \ref{CAF} unavailable. We can however say some general things about the structure of the beta function, which now takes the form $\beta_{\alpha} = -B' \alpha^2 + C' \alpha^3 + D \alpha^4$, where $B'$ and $C'$, unlike the case of the 2-loop beta function, depend on the quartic couplings. 
Close to the Gaussian fixed point, following the reasoning of Sec.~\ref{CAF}, the quartic couplings scale as $\lambda_{i}\propto \alpha \ll 1$, and the new terms in $B'$ and $C'$  become of order $\alpha^4$, such that $B'$ and $C'$ in this limit become  $B$ and $C$ from the 2-loop case. 

Similarly, we know from App.~\ref{app:UVIR} that $\lambda_{i}^* = \lambda_{i s}^* \alpha^*$, so as long as $ \lambda_{i s}^* \sim \mathcal{O}(1)$ and $\alpha^* \ll 1$, we obtain the same result that the dependence of the quartic couplings is moved to the $D$ coefficient. 

Here we will first argue that $\lambda_{i s}^* \sim \mathcal{O}(1)$ or smaller, and then use this result to show that the three-loop contribution to $\alpha^*$ is sub leading in certain limits. Afterwards we will use a numerical approach to determine the IR fixed points and produce a table similar to Table \ref{fpwindow}. 

Studying the rescaled one-loop beta functions (left hand side of Eq.~(\ref{eq:pseudo}) in the large $N_c$ and $N_s$, but fixed $N_s / N_c = k$, limit, we find the rescaled couplings to be roughly of the order (leading term)
\begin{align}
\mathrm{FP}_1 &= \left\{\frac{3-\sqrt{9-3 k}}{4 k }\frac{1}{N_c},\frac{3-\sqrt{6-3 k}}{4 (k+1)}\right\}\nonumber \\
\mathrm{FP}_2 &= \left\{\frac{3+\sqrt{9-3 k}}{4 k }\frac{1}{N_c},\frac{3-\sqrt{6-3 k}}{4 (k+1)}\right\}\nonumber \\
\mathrm{FP}_3 &= \mathrm{FP}_4 = \left\{-\frac{\sqrt{3} (k+1) \sqrt{k \left(k^2-k+10\right)}+9 k}{4 \left(k^3-k^2+k\right)}\frac{1}{N_c},\frac{3+\sqrt{6-3 k}}{4 (k+1)}\right\}\, ,
\end{align}
where format is $\mathrm{FP}_i = \left\{\lambda_{1s},\lambda_{2s}\right\}$, and FP$_1$, FP$_2$ are the two fixed points we have in regions with two solutions, while FP$_3$, FP$_4$ are the additional ones in the four solution region. The value of $k$ is constrained from the slopes of the borders of the grey regions on Fig.~\ref{Region2}, which means $k\in\left[0,  0.84 \right]$. Notice, they all are of the order $\left\{\mathcal{O}\left(1/N_c\right), \mathcal{O}\left(1\right)\right\}$. 

The fixed point at two loop order is given by 
\beq\label{alpha2}
\alpha^*_2 = \frac{B}{C} =  \frac{8 N_x N_c}{150 N_c^3+N_c^2 \left(3 N_s-52 N_x\right)-66 N_c-9 N_s+12 N_x}\, ,
\eeq
which in the same limit as taken above is
\beq\label{alphapert}
\alpha^*_2 = \frac{8 }{3 (k+50)}\,\frac{N_x}{ N_c^2} + \mathcal{O}\left(\frac{N_x^2}{N_c^3}\right)
\eeq

\begin{figure}[t]
	\centering
	\includegraphics[width=0.45\columnwidth]{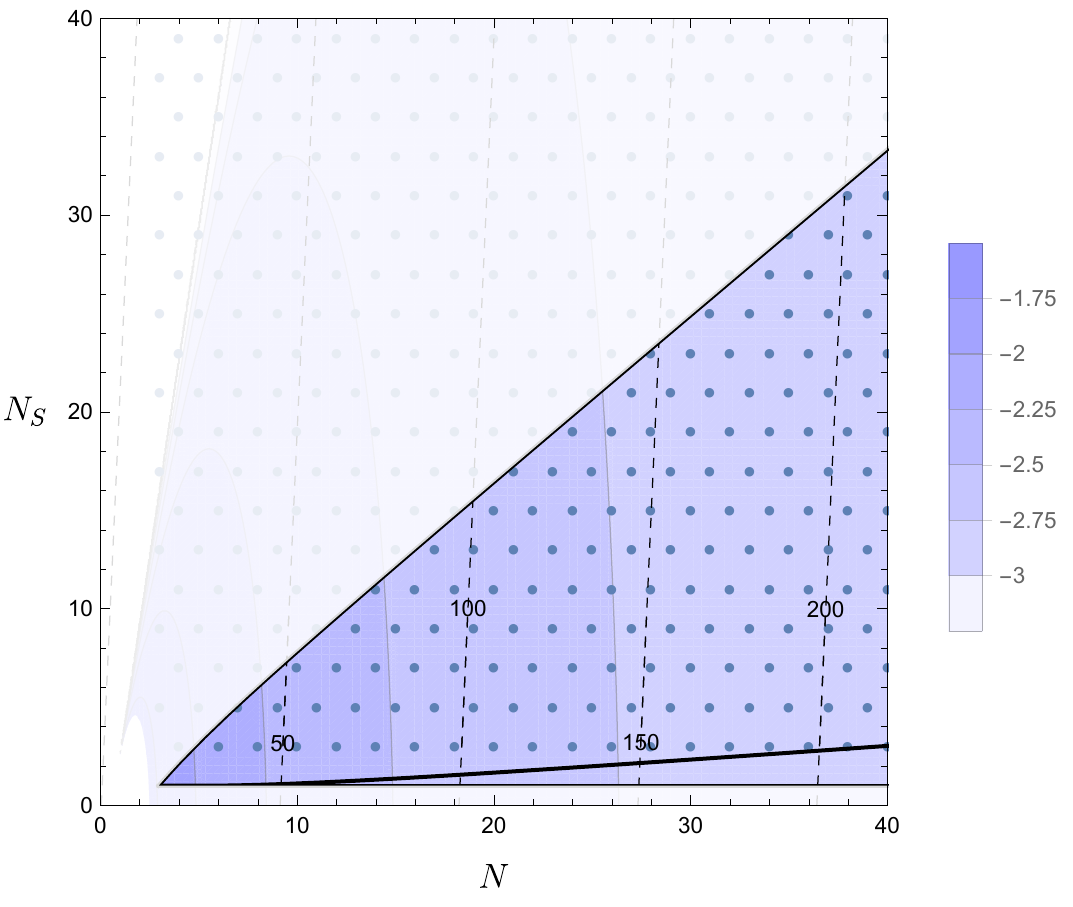}
	\hspace{0.05\columnwidth}
	\includegraphics[width=0.45\columnwidth]{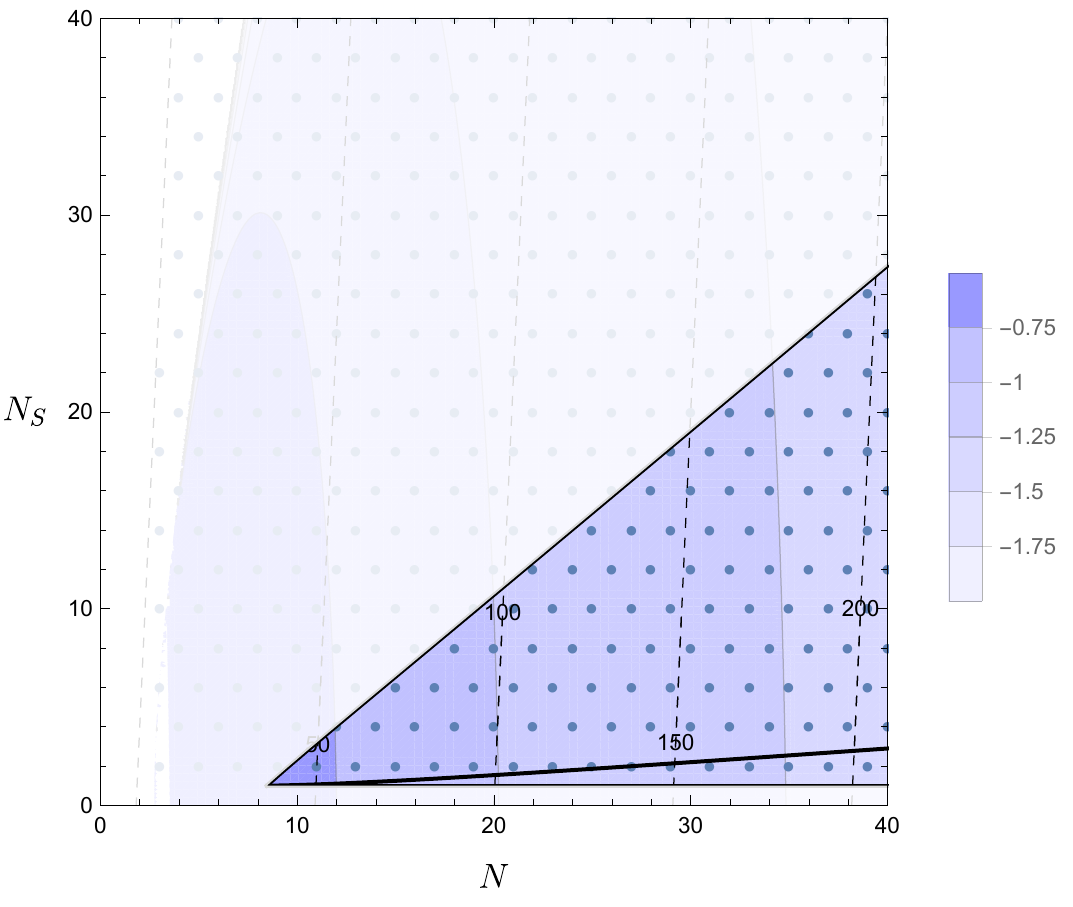}
	\caption{Estimate of the relative size of the three-loop contribution. The dots mark integer values of $N_c$, $N_s$ and $N_f$ such that $N_x = 1/4$ in the left panel and $N_x = 10$ on the right. The excluded regions does not satisfy the CAF conditions. The dashed contours are values of $N_f$, while colored regions show the relatica size of the three-loop contribution as $\log_{10} \left| \alpha^*_{3}-\alpha^*_{2}\right| / \alpha^*_{2}$, with $\alpha^*_{2}$ given by Eq.~(\ref{alpha2}), while $\alpha^*_{3}$ is the three loop result assuming for simplicity that $\lambda_{1} = \lambda_{2} = \alpha$.}
	\label{SizeofThree}
\end{figure}

With these results at hand, we can determine the dominating terms in Eq.~(\ref{betag3}). There are $N^3$-terms (counting $N_c$, $N_s$ and $N_f$ as $N$) coming from both $B'$, $C'$ and $D$, where the contributions from $B'$ ($C'$) depend quadratically (linearly) on $\lambda_{2s}$. However, when $\alpha^* \propto N_x / N_c^2$, these terms are suppressed by a factor of $N_x/N_c$ and will thus only contribute to the sub leading terms of Eq.~(\ref{alphapert}). The degree of suppression can roughly be estimated by comparing the blue solid line with the dashed red line in Fig.~\ref{fig:NNx}. Assuming instead $\lambda_{1s} =\lambda_{2s}= 1$, we can easily calculate the size of the corrections. These are shown in Fig.~\ref{SizeofThree} for $N_x = 1/4$ and $N_x = 10$, respectively. 

Following the numerical approach to determine the existence of IR fixed points, by solving  $(\beta_{\alpha}, \beta_1, \beta_2) = 0$, we the results in Table~\ref{fpwindow3loop}.
\begin{table}[ht]
	\centering
	\resizebox{\linewidth}{!}{%
		\begin{tabular}{ | c | c | c | c | c | c | c | c | c | c |} 
			\hline
			& $N_c = 4$ & $N_c = 5$ & $N_c = 6$ & $N_c = 7$ & $N_c = 8$ & $N_c = 9$ & $N_c = 10$ & $N_c = 11$ & $N_c = 12$ \\ \hline
			$N_s = 2$ & $10-21$ & $10-26$ & $11-32$ & $13-37$ & $15-43$ & $17-48$ & $19-54$ & $21-59$ & $23-65$ \\ \hline
			$N_s = 3$ & & $10-26$ & $11-32$ & $13-37$ & $15-43$ & $17-48$ & $19-54$ & $21-59$ & $22-65$ \\ \hline
			$N_s = 4$ & & & $11-31$ & $13-37$ & $14-42$ & $16-48$ & $18-53$ & $20-59$ & $22-64$ \\ \hline
			$N_s = 5$ & & & & $12-37$ & $14-42$ & $16-48$ & $18-53$ & $20-59$ & $22-64$ \\ \hline
			$N_s = 6$ & & & & & $14-42$ & $16-47$ & $17-53$ & $19-58$ & $21-64$ \\ \hline
			$N_s = 7$ & & & & & & $15-47$ & $17-53$ & $19-58$ & $21-64$ \\ \hline
			$N_s = 8$ & & & & & & & & $19-58$ & $21-63$ \\ \hline
			$N_s = 9$ & & & & & & & & & $20-63$ \\ \hline
		\end{tabular}}
		\caption{Window in $N_f$ that allow for perturbative IR fixed points for $N_s = \{2,9\}$, $N_c = \{4,12\}$, when including the three loop contributions of the gauge coupling beta function. There are no solutions for $N_c = 3$ for $N_s >1$.}
		\label{fpwindow3loop}
	\end{table}
	
	As expected, when comparing Table \ref{fpwindow} and \ref{fpwindow3loop} we see that the upper boundary of the window in $N_f$ remains unchanged when including the 3-loop contribution to the gauge beta function. The lower boundary, however, is lowered. Furthermore, we find the existence of another lower limit of $N_f$ which lies above the CAF solutions. For any value of $N_f$ above this limit the system exhibits IR fixed points. These are not shown in the table.

\section{Diagonalization of S}\label{diagonalisation}
  We start by left-multiplying $S$ by an appropriate $SU(N_c)$ matrix.
  To find it, note that an arbitrary unitary matrix satisfies the following two properties:
  \begin{itemize}
    \item Each row i satisfies $\sum_j |U_{ij}|^2 = 1$. This imposes a constraint on one degree of freedom for each row.
    \item Each pair of columns ($i$,$j$) with $i\neq j$ satisfies $\sum_k U_{ki}^*U_{kj} = 0$. Each of these imposes a constraint on two degrees of freedom (one complex number).
  \end{itemize}
  Proceeding row-by-row, the first row has $2N_c - 1$ degrees of freedom (only the first constraint) and the $n$th row now has $2N_c - 2n + 1$ degrees of freedom (first constraint and $n-1$ second constraints).
  Now, this matrix multiples the $N_c \times N_s$ complex matrix. We can conclude that
  \begin{itemize}
    \item If $2N_c - 2n + 1 \geq 2N_s$ we have enough freedom to set all entries of the $n$-th row of $S$ to $0$. Solving for $n$ we can conclude that we can set $N_c - N_s$ rows to $0$.
    \item For row numbers greater than $N_c - N_s$ we will have $2N_s - 2N_c +2n -1$ degrees of freedom remaining. Starting from $n = N_c - N_s + 1$ we will be left with $1,3,5,\ldots$ degrees of freedom, which translates to one real number and $0,1,2,\ldots$ complex numbers.
  \end{itemize}
  Having cast $S$ into a triangular form, we can right-multiply it by an $SU(N_s)$ matrix and repeat the argument to arrive with matrix that is diagonal in one $N_s\times N_s$ block.

 Decomposing $D$ in terms of a vacuum expectation value part ($\Sigma$) and a diagonal perturbation ($H$), the field $S$ can finally be written in the form:
  \begin{equation}
    S = e^{i \pi_c^a(x) t_c^a/f_c} \left( \Sigma + H(x) \right) e^{-i \pi_s^a(x) t_s^a/f_s}
    \label{eq:non-linear-sigma1}
  \end{equation}
  From the $\Sigma$ term we can find the combination of generators which do not leave the vacuum invariant, which will define our symmetry breaking pattern.
  These generators will correspond to the Goldstone bosons.
  Note that the Goldstone fields corresponding to the colour symmetry, $\pi_c$, can be set to zero by an appropriate choice of gauge (unitary gauge), in which some of the vector bosons become massive. 
  The Goldstone fields corresponding to flavour symmetry on the other hand become real, massless degrees of freedom.\\

\section{Symmetry Breaking Patterns}\label{app:symmetrypatterns}

In Sec.~\ref{tree-analysis}, we found from the tree-level analysis two possible vacuum configurations of the scalar fields. 

For $\lambda_2 < 0$, the tree-level potential is minimized when there is only one non-zero vacuum expectations value of the scalar matrix, $S$, i.e.  
\begin{equation}
	\langle S_{ia} \rangle = \rho \delta_{i1} \delta_{a1}.
\end{equation}

For $\lambda_2 > 0$, the potential is minimized when there is $N_s$ non-zero elements, i.e. 
\begin{equation}
	\langle S_{ia} \rangle = \rho \delta_{ia}.
\end{equation}

 In the following we will derive the corresponding symmetry breaking patterns. To do so, we return to eq. \ref{eq:non-linear-sigma1} and note first that canonical normalisation of $\pi$ fields requires $f_c=f_s=\sqrt{\rho}$ in both symmetry breaking cases. We are looking for combinations of $\pi^a_c$ and $\pi^a_s$, which leave the vacuum state invariant, in other words satisfy the relation
 \begin{equation}
 \pi^a_c(x) t_c^a \Sigma - \pi_s^b \Sigma t_s^b = 0.  
 \label{eq:sb_cond}
 \end{equation}
 We will now consider both cases separately, starting with the $\Sigma_{ia} = \rho \delta_{ia}$ case.
 Assuming $N_c>N_s$, we can divide the total of $N_s^2 + N_c^2 - 1$ generators (including $N_c+N_s-1$ diagonal ones) of $SU(N_c)\times U(N_s)$ into four different categories:
 \begin{enumerate}
 	\item All colour generators, with non-zero entries when both indices are in range $N_s+1,\ldots,N_c$ (total: $(N_c-N_s)^2-1$ out of which $N_c-N_s-1$ are diagonal).
 	These clearly satisfy \ref{eq:sb_cond} with $\pi_s^a = 0$. These generators form an $SU(N_c - N_s)$ algebra.
 	\item All colour generators with non-zero entries when both indices are in $1,\ldots,N_s$ range and all flavour generators except identity (total $2N_s^2-2$ out of which $2 N_s-2$ are diagonal).
 	In this case we can choose the generators such that $t_c^a = t_s^a = t^a$, and noting that all $t^a$ commute with $\Sigma$ (because it is a diagonal matrix), we can write \ref{eq:sb_cond} as
 	\begin{equation}
 	(\pi^a_c-\pi^a_s) t^a = 0,
 	\end{equation}
 	which implies that $\pi^a_c = \pi^a_s$ by linear independence of SU(N) generators.
 	This implies that there are $N_s^2 - 1$ unbroken generators when $\pi^a_c = \pi^a_s$, which form an algebra of $SU(N_s)$ and $N_s^2 - 1$ broken generators.
 	\item Remaining diagonal generators (total: 2).
 	These are the identity matrix of $U(N_s)$ and one matrix from the Cartan subalgebra of $SU(N_c)$, which we will call $t^0$.
 	We require $\mathrm{Tr} t^0 t^a_c = 0$ for all $a\neq0$ and in particular for all the diagonal Cartan subalgebra generators. But we have already considered above all the generators which form an $SU(N_s)$ Cartan subalgebra on the first $N_s$ entries and are zero on the remaining ones and all the generators that form a Cartan subalgebra on the last $N_c-N_s$ entries and are zero in the first $N_s$ entries.
 	The only way the trace condition can be satisfied is if $t^0$ is separately proportional to the identity matrix on the first $N_s$ entries and on the last $N_c-N_s$ entries, with proportionality constants chosen in such a way that it is traceless (as required by the algebra of $SU(N_c)$).
 	But the fact that the matrix is proportional to the identity on the first $N_s$ entries implies that $t^0 \Sigma \propto \Sigma$ and we can choose the coefficient $\pi^0_c$ together with the coefficient of the identity matrix in flavour space in such a way that the symmetry is unbroken.
 	As a consequence, we have one unbroken generator, which corresponds to the $U(1)$ symmetry, while the other generator is broken.
 	\item The next type of generator we will consider are off-diagonal generators with non-zero entries where one index is in ($1\ldots N_s$) range and the other is in ($N_s+1,\ldots N_c$) range ($2(N_c-N_s)N_s$ total).
 	These will necessarily break the vacuum, because the second term of eq. \ref{eq:sb_cond} would need to transform a row of zeroes into one containing a $\rho$.
 \end{enumerate}
 We have thus classified all the generators of $SU(N_c)\times U(N_s)$ and conclude that the symmetry breaking pattern corresponding to vacuum 1 is
 \begin{equation}
 SU(N_c)\times U(N_s) \to SU(N_c-N_s) \times SU(N_s) \times U(1)\, .
 \end{equation}
 
 The other symmetry breaking pattern has the vacuum configuration which is zero everywhere except for the (1,1) entry.
 We again discuss the broken and unbroken generators by splitting $N_c^2 + N_s^2 - 1$ generators into three different groups.
 \begin{enumerate}
 	\item All generators of $SU(N_c)$ and $U(N_s)$ with vanishing first row and column ($(N_c - 1)^2 + (N_s-1)^2 - 1$ generators, including $N_c + N_s - 3$ diagonal ones)
 	These generators annihilate the vacuum by themselves and are therefore all unbroken. The corresponding symmetry subgroup is $SU(N_c - 1) \times U(N_s-1)$
 	\item All off-diagonal generators with non-vanishing entries in the first row and column ($2(N_c+N_s -2)$ generators)
 	The generators of $SU(N_c)$ will change the row in which $\rho$ appears, while the generators of $SU(N_s)$ will change the column in which the $\rho$ appears, so there is no combination, which will leave the vacuum invariant - all these generators are broken.
 	\item Remaining diagonal generators (2 generators)
 	The remaining generators are one $SU(N_c)$ generator and one $SU(N_s)$ generator, which are diagonal with non-zero first entry.
 	This implies that the action of each of these generators on the vacuum will be proportional to the vacuum state, and we can tune the corresponding fields so that the symmetry generator corresponding to this linear combination vanishes, as before. We are therefore left with one broken generator and one unbroken generator, which corresponds to anouther $U(1)$ symmetry.
 \end{enumerate}
 In conclusion, the symmetry breaking pattern in this case is
 \begin{equation}
 SU(N_c)\times U(N_s) \to SU(N_c-1) \times U(N_s-1) \times U(1)\, .
 \end{equation}

\section{Renormalization group improved effective potential}\label{app:RGimproved}
The formalism for studying spontaneous symmetry breakdown using the RG improved effective potential was introduced in Sec.~\ref{loop level eff-analysis}. Here we will analyze the derived conditions for the case with $\left(N_c=6, N_s=3, N_f=31\right)$.

\begin{figure}[htb]
	\centering
	\includegraphics[width=0.6\columnwidth]{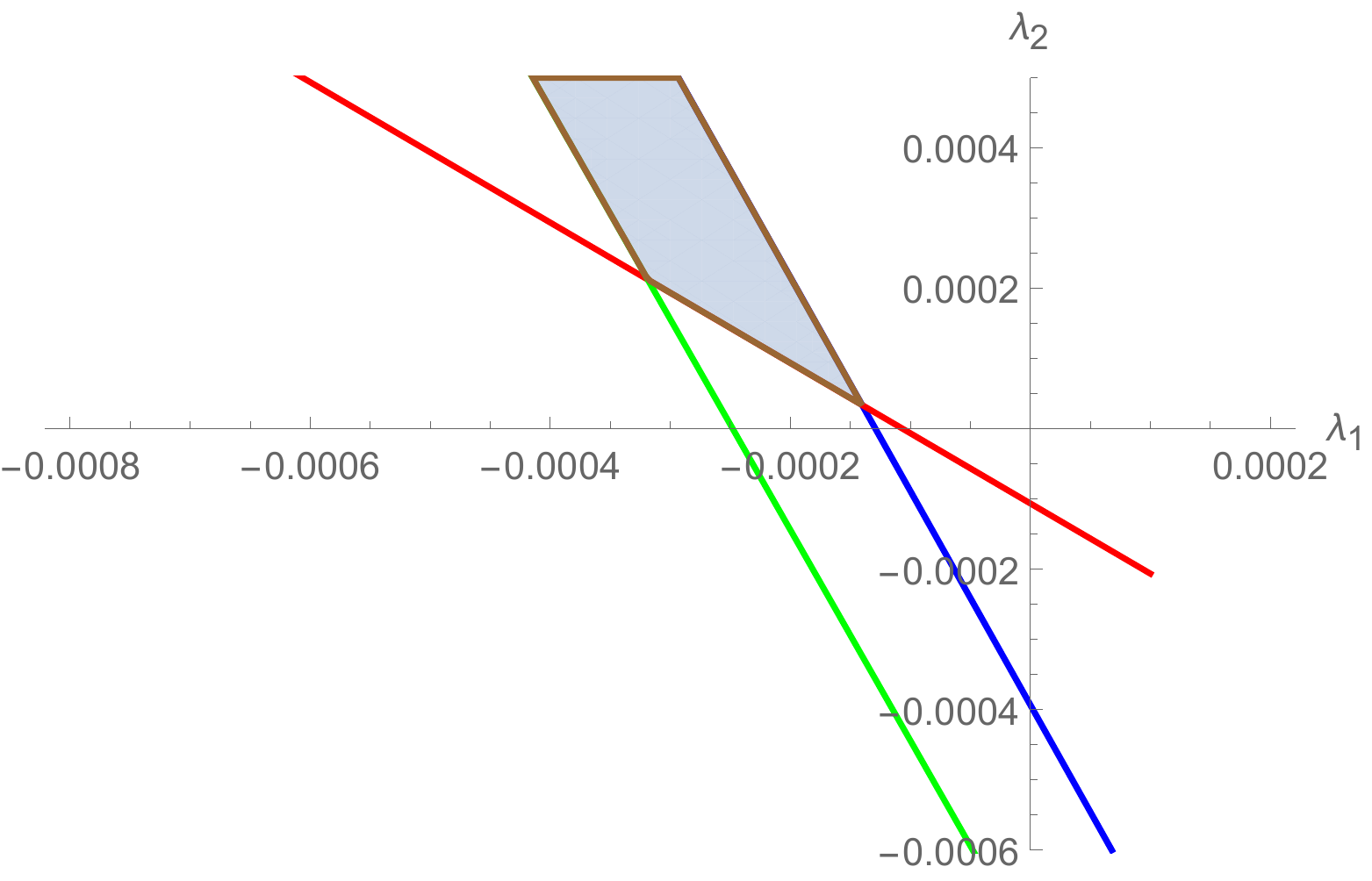}\hspace{0.05\columnwidth}
	\caption{In this figure we choose a particular slice at $g=0.1$ perpendicular to the gauge coupling direction. The blue line corresponds to the symmetry breaking boundary line while the green and red lines come from two vacuum stability conditions. The blue shaded region represent the broken phase $SU(N_c)\times U(N_s) \to SU(N_c-N_s) \times SU(N_s) \times U(1)$.}
	\label{Region Plot_1}
\end{figure}

We first focus on the case where $\kappa=1$, which corresponds to the case of eq.~\eqref{SB pattern 2} and plot the results in figure~\ref{Region Plot_1}.
We calculate the Hessian matrix using Eq.~\eqref{Hessian matrix}. The first two eigenvalues are degenerate and lead to the following constraint:
\begin{equation}
\lambda _2\geq\frac{1}{16} \left(-17 g^4-16 \lambda _1\right)\,,\label{vacuum stable 1}
\end{equation}
where we have already ignored the higher order terms $\lambda_i^n\,\left(n\geq 2\right)$ and $g^m\,\left(m\geq 4\right)$ in the above expression. 
The above vacuum stable line is shown in red in figure~\ref{Region Plot_1}. 
The third mass eigenvalue will lead to (again ignoring higher order terms)
\begin{equation}
\lambda _2\geq\frac{1}{16} \left(-119 g^4-48 \lambda _1\right)
\label{vacuum stable 2}
\end{equation}
which is shown in green in figure~\ref{Region Plot_1}. 
In addition, the RG improved boundary line will be given by eq.~\eqref{RG improved VEV condition} with respect to $k=1$, leading to
\begin{equation}
35 g^2-240 \lambda _1-2+\sqrt{-611 g^4+13440 g^2 \lambda _1-140 g^2+38592 \lambda _1^2-768 \lambda _1+4}=0 \label{RG improved boundary 1}\,,
\end{equation}
which is the blue line in figure~\ref{Region Plot_1}.

We combine the above two vacuum stability lines with the RG improved boundary line for the broken phase in  figure~\ref{Region Plot_1}. 
It is very clear that when RG flows run into the shaded region (shown in blue), the symmetry is broken and we have the symmetry breaking pattern:  $SU(N_c)\times U(N_s) \to SU(N_c-N_s) \times SU(N_s) \times U(1)$.

\begin{figure}[htb]
	\centering
	\includegraphics[width=0.6\columnwidth]{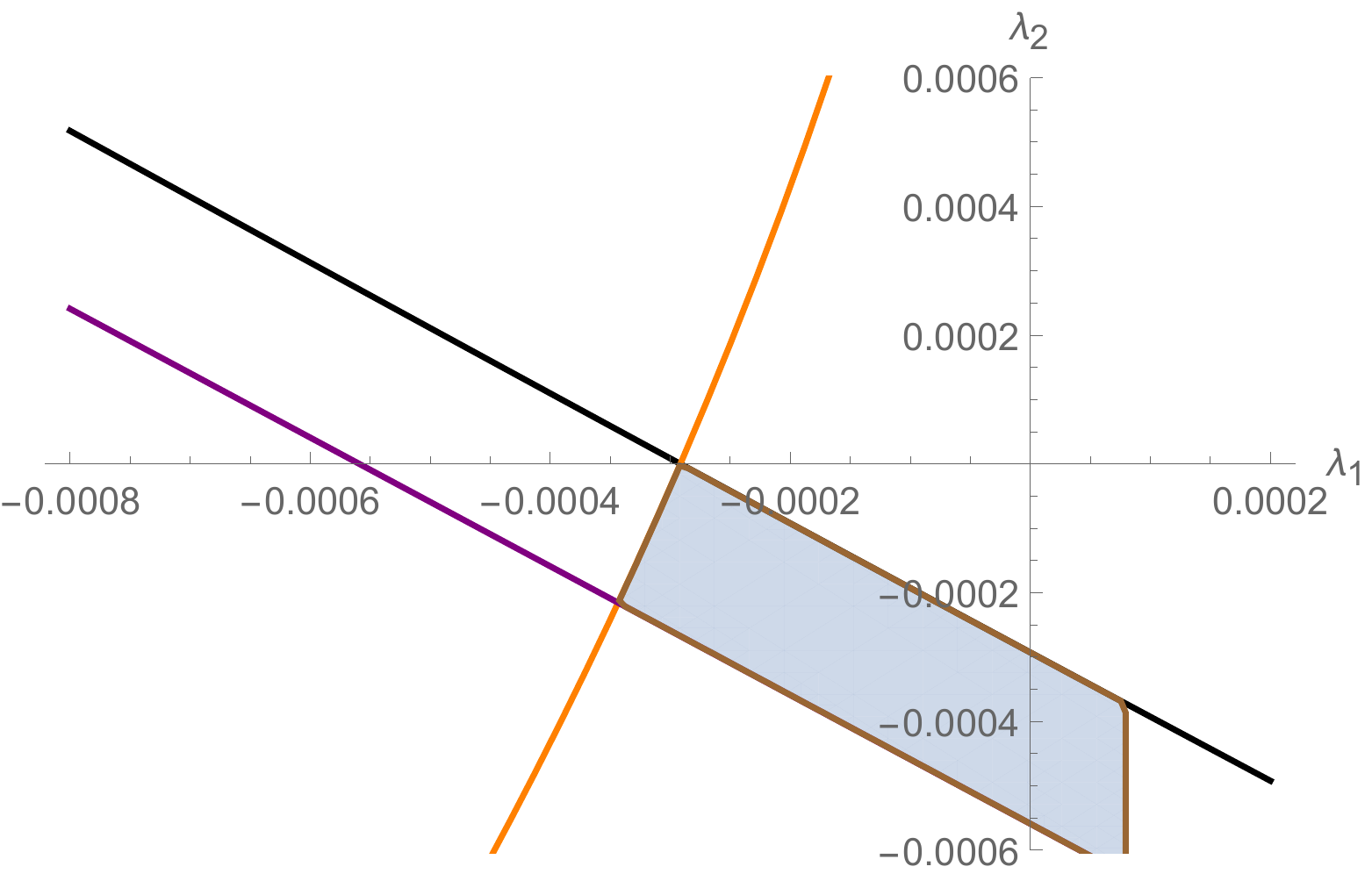}\hspace{0.05\columnwidth}
	\caption{In this figure we choose a particular slice at $g=0.1$ perpendicular to the gauge coupling direction. The black line corresponds to the symmetry breaking boundary line while the orange and purple lines come from two vacumm stability conditions.
		The blue shaded region represent the broken phase $SU(N_c) \times U(N_s) \rightarrow SU(N_c-1) \times U(N_s -1) \times U(1)$.}
	\label{Region Plot_2}
\end{figure}

We likewise plot in figure~\ref{Region Plot_2} the results for $\kappa=0$. In this case the two degenerate mass eigenvalues and one non-degenerate mass eigenvalue provide the following two constraints
\begin{equation}
\lambda _2\geq\frac{1}{144} \left(-805 g^4-144 \lambda _1\right);\quad \lambda _2\leq\frac{1}{96} \left(35 g^2-96 \lambda _1-\sqrt{305 g^4+4992 \lambda _1^2-384 \lambda _1}\right)\,,
\label{vacuum stable 3}
\end{equation}
which correspond to the purple and orange lines respectively in figure~\ref{Region Plot_2}. 
Furthermore,  by using eq.~\eqref{RG improved VEV condition} with respect to $\kappa=0$, the RG improved boundary line for the broken phase ($\kappa=0$) is
\begin{equation}
\lambda _2\leq\frac{1}{96} \left(35 g^2-96 \lambda _1-2\sqrt{305 g^4-140 g^2+4992 \lambda _1^2+4}\right)
\label{RG improved boundary 2}
\end{equation}
and is shown in black in figure~\ref{Region Plot_2}. The  two vacuum stability lines and the RG improved boundary line for the broken phase (for $\kappa=0$ case) are together illustrated in  figure~\ref{Region Plot_2}.

\begin{figure}[htb]
	\centering
	\includegraphics[width=0.6\columnwidth]{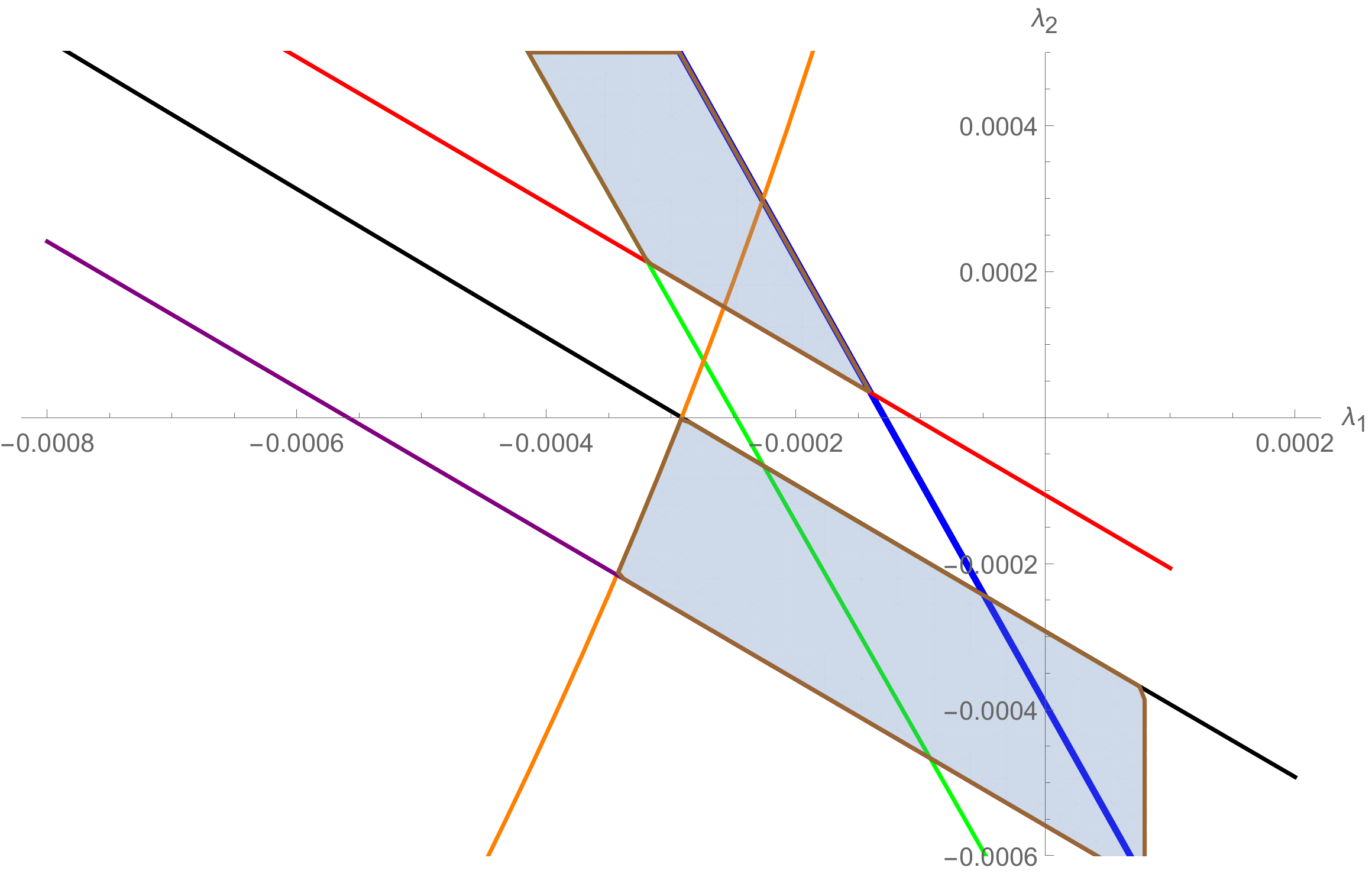}\hspace{0.05\columnwidth}
	\caption{In this figure we choose a particular slice at $g=0.1$ perpendicular to the gauge coupling direction. The blue and black lines correspond to the symmetry breaking boundary lines while the orange, purple, red and green lines come from four vacumm stability conditions respectively.
		The two shaded regions represent the broken phases $SU(N_c)\times U(N_s) \to SU(N_c-N_s) \times SU(N_s) \times U(1)$ and $SU(N_c) \times U(N_s) \rightarrow SU(N_c-1) \times U(N_s -1) \times U(1)$ respectively.}
	\label{Region Plot_3}
\end{figure}

In figure~\ref{Region Plot_3} we combine the above two cases, with the two shaded regions representing the broken phases $SU(N_c)\times U(N_s) \to SU(N_c-N_s) \times SU(N_s) \times U(1)$ and $SU(N_c) \times U(N_s) \rightarrow SU(N_c-1) \times U(N_s -1) \times U(1)$ respectively. 
Note that figure~\ref{Region Plot_3} is consistent with the tree level diagram (figure~\ref{tree level region plot}) in the previous section. 
From the previous section it is clear that the RG improved boundary lines  \eqref{RG improved boundary 1} and \eqref{RG improved boundary 2}  for the broken phases actually shift the tree level boundary lines slightly, as described in  \eqref{broken phase boundary line_1} and \eqref{broken phase boundary line_2}, at the origin of the coupling space $(\lambda_1,\lambda_2)$. 
Magnification near the origin  of figure~\ref{tree level region plot} yields the detailed structure shown in figure~\ref{Region Plot_3}. 
When scalar couplings are large  the coarse grained picture   in figure~\ref{tree level region plot} emerges.

\begin{figure}[htb]
	\centering
	\includegraphics[width=0.6\columnwidth]{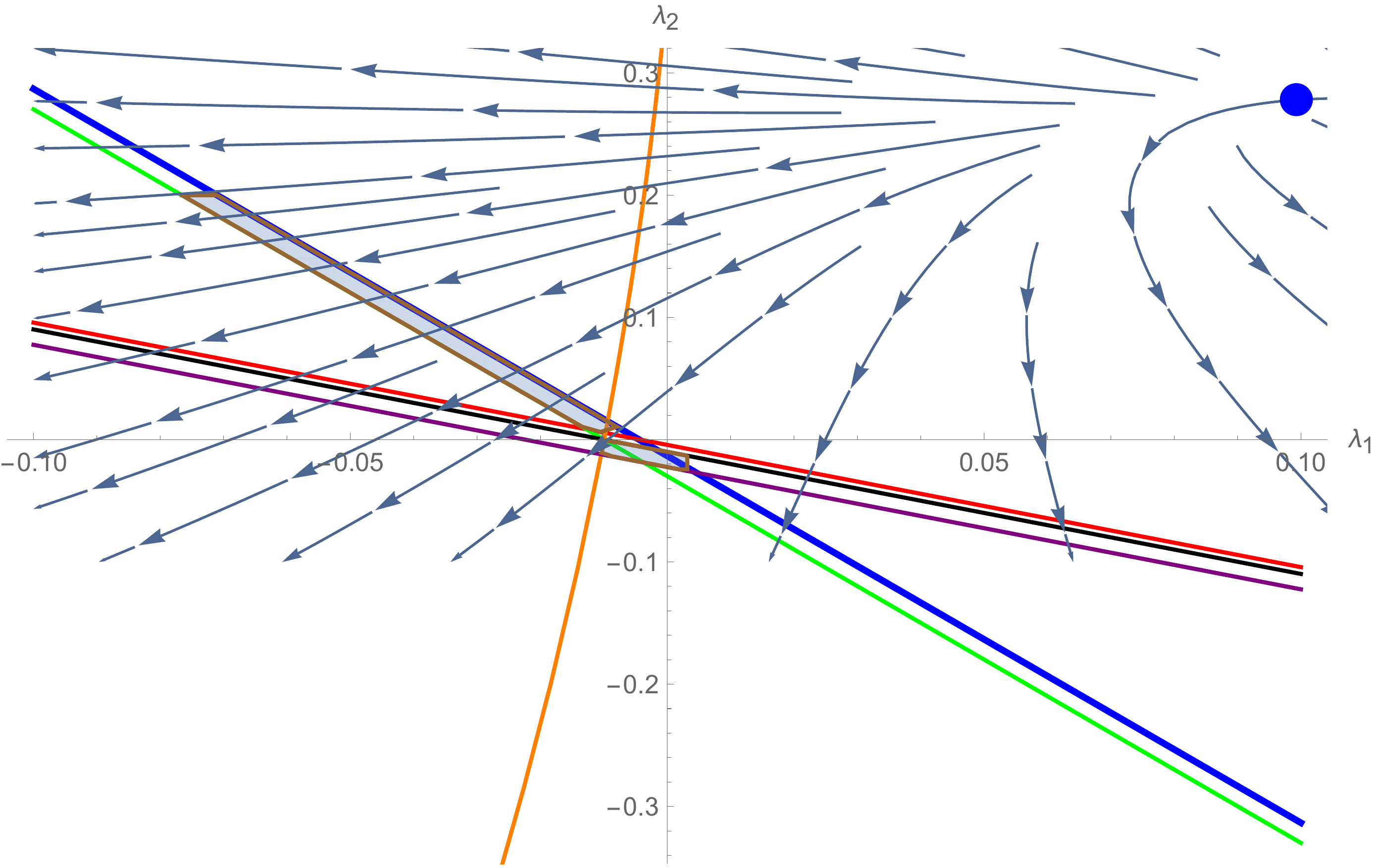}\hspace{0.05\columnwidth}
	\caption{In this figure we choose a particular slice $g=0.044$ which is the coupling value of the Banks-Zaks fixed points. The two shaded regions represent the broken phases $SU(N_c) \times U(N_s) \rightarrow SU(N_s) \times SU(N_c - N_s)\times U(1)$ and $SU(N_c) \times U(N_s) \rightarrow SU(N_c-1) \times U(N_s -1) \times U(1)$ respectively.}
	\label{Region Plot_4}
\end{figure}

Combining the boundary lines for the broken phases and the vacuum stability lines with the RG flows yields figure~\ref{Region Plot_4}. We have chosen the 
particular slice at $g=0.044$ which is the coupling value of the Banks-Zaks fixed points.
Note that the scalar quartic couplings have been rescaled to be compatible with the stream plot;  the couplings on the axes are the rescaled couplings (with much larger values than the physical couplings). 

The Banks Zaks fixed point shown in blue is fully repulsive in this slice and plays the role of an interacting UV fixed point (in this 2D slice). It is very clear that there are RG flows running from this interacting UV fixed point (blue) towards the CP3 region (where $\lambda\sim O\left(g^n\right)\,\left(n\geq4\right)$, shown in shaded blue). It is also clear that in the 2D slice there are also RG flows running from interacting UV fixed point towards the Gildener Weinberg region (where $\lambda\sim O\left(g^2\right)$) of the broken phase.
Since there is no particular boundary between these two regions, we obtain phases with complete asymptotic safety in the UV (with perturbative couplings) and symmetry breaking in the IR regardless of whether scalar couplings scale with lower powers of the gauge coupling (Gildener-Weinberg) or higher  (CP3).  Furthermore, there are two attractive directions of the fixed point (one coming from the UV Gaussian fixed point and run into the plane (through back of the plane) the other one running into the plane through the front). Hence, there are some flows that do not come directly from the blue fixed point in this 2D slice but run from the UV Gaussian fixed point, passing the IR fixed point and run towards the CP3 region. A similar conclusion holds for the Gildener Weinberg region.
Thus there exist phases that are completely asymptotically free in the UV, symmetry breaking in the IR, and walking behaviour in the middle regardless of whether scalar couplings scale with lower powers of the gauge coupling (Gildener-Weinberg) or higher  (CP3).

\begin{table}[ht]
	\centering
	\begin{tabular}{|| l | l | l ||}
		\hline
		Scenarios & Boundary Lines & Symmetry Breaking Pattern \\ \hline (tree) $\lambda_2>0$ & eq.~\eqref{broken phase boundary line_1} & $SU(N_c) \times U(N_s) \rightarrow SU(N_s) \times SU(N_c - N_s)\times U(1)$ \\ \hline (tree) $\lambda_2<0$ & eq.~\eqref{broken phase boundary line_2} & $SU(N_c) \times U(N_s) \rightarrow SU(N_c-1) \times U(N_s -1) \times U(1)$ \\ \hline (loop) $\lambda_2>0$ & eq.~\eqref{RG improved boundary 1} & $SU(N_c) \times U(N_s) \rightarrow SU(N_s) \times SU(N_c - N_s)\times U(1)$ \\ \hline (loop) $\lambda_2<0$ & eq.~\eqref{RG improved boundary 2} & $SU(N_c) \times U(N_s) \rightarrow SU(N_c-1) \times U(N_s -1) \times U(1)$\\ \hline
	\end{tabular}
	\caption{Two categories (tree level and loop level analysis) and four scenarios (according to whether $\lambda_2>0$ or $\lambda_2<0$) are summarized in the table.}
	\label{summary_1}
\end{table}

\section{One Loop Effective Potential}\label{eff-explicit}

In section \eqref{loop level eff-analysis} we used  RG improvement  to analyze the effective potential. This approach has several advantages:
\begin{itemize}
  \item Loop level contributions are already encoded in the RG functions. No explicit calculations of  loop contributions to the effective potential are required.
  \item No initial assumption about $\lambda\sim g^4$ is required. All orders of $g$ are summed and already encoded.
  \item Both gauge loop  and scalar loop contributions are included.
  \item It is much easier to  generalize this approach to different symmetry groups and representations.
\end{itemize}

In this appendix we  sketch an explicit way to analyze the effective potential. To simplify the calculation, we study the $SU(3_c)\times U(3_s)$ case and assume the tree level contribution $O\left(\lambda\right)$  is comparable to the one loop   gauge contribution $g^4$ (i.e.~$\lambda\sim g^4$), implying the next-order scalar contributions can be ignored ($\lambda^2\sim g^8$).  Explicit calculations are carried out in the Coleman-Weinberg scheme, which satisfies the Coleman-Weinberg renormalization conditions (discussed below) \cite{Coleman:1973jx,Gildener:1975cj}.

The one loop effective potential is given by
\begin{equation}
V_{eff}^{1loop}=V_0+V_g+V_{ct} 
\end{equation}
where $V_0, V_g, V_{ct}$ represent the tree level term, the gauge loop contribution and the counter-terms respectively. The one loop gauge contribution can be further written as \cite{Jackiw:1974cv}
\begin{equation}
V_g=\frac{3}{64\pi^2}\Tr\left[M^4\left(\boldsymbol{S}\right)\log M^2\left(\boldsymbol{S}\right)\right]\,,
\end{equation}
where $\boldsymbol{S}$ is the scalar field under fundamental representation of $SU(3_c)\times U(3_s)$ and
\begin{equation}
M_{ab}^2=g^2t_{ji}^at_{ik}^b S_j^{f\dagger}S_k^{f}=g^2t_{ji}^at_{ik}^b\boldsymbol{\chi_{ji}}\,\quad \boldsymbol{\chi_{ji}}=
\textrm{diag}\left(\rho_1^2,\rho_2^2,\rho_3^2\right)\quad\left(a,b=1,\cdots8\right)\,.
\end{equation}
The diagonalization of  $\boldsymbol{S}$ is discussed in great detail below eq.~\eqref{diagonalize}. Using this, the one loop effective potential can be explicitly written as
\begin{equation}
\begin{split}
V=&\left(a_1+\lambda _1\right) \left(\sum _{i=1}^L \rho _i^2\right)^2+\left(a_2+\lambda _2\right) \sum _{i=1}^L \rho _i^4+\frac{3 g^4}{64 \pi ^2}\bigg(2 \sum _{i=1}^L \sum _{j=1}^{i-1} \left(\rho _i^2+\rho _j^2\right)^2 \log \left(\rho _i^2+\rho _j^2\right) \\
\qquad +&M_-^2 \log \left(M_-\right)+M_+^2 \log \left(M_+\right)\bigg)\,, \quad \left(L=3\right)
\label{effective potential explicit}
\end{split}
\end{equation}
where $\rho_i^2+\rho_j^2\,\left(i,j=1,2,3\right)$ are the six polynomial eigenvalues of the mass matrix $M_{ab}^2$ while $M_+,\,M_-$ are the two non-polynomial eigenvalues, written explicitly as
\begin{equation}
M_{\pm}=\frac{2}{3} \left(\rho _1^2+\rho _2^2+\rho _3^2\pm\sqrt{\rho _1^4-\rho _2^2 \rho _1^2-\rho _3^2 \rho _1^2+\rho _2^4+\rho _3^4-\rho _2^2 \rho _3^2}\right)\,,
\end{equation}
and $a_1,\,a_2$ are the counter-terms which are determined through the Coleman-Weinberg renormazliation conditions. 

The Coleman-Weinberg conditions are  
\begin{equation}
\begin{split}
\frac{1}{4N_c\left(N_c-1\right)}\sum_{i-1}^{N_c}\sum_{j=1}^{i-1}\frac{\partial^4V}{\partial\rho_i^2\partial\rho_j^2}\bigg|_{\rho_i=\kappa_iM_R}
&=\lambda_1\qquad \left(N_c\geq N_s\right)\\
\frac{1}{4!N_c}\sum_{i=1}^{N_c}\frac{\partial^4V}{\partial\rho_i^4}\bigg|_{\rho_i=\kappa_iM_R}
&=\lambda_1+\lambda_2\,,\label{CW conditions_appendix}
\end{split}
\end{equation}
where $M_R$ is the renormalization scale and $\kappa_i$ represents the relative ratio between different scales. Using  eq.\eqref{CW conditions_appendix} we can determine $a_1,\,a_2$. Inserting the result into eq.~\eqref{effective potential explicit}, we further obtain the full expression for the  effective potential $V_{eff}^{1loop}$. This expression is extremely long and not particularly illuminating, so we do not present it.

The next step is to study the one loop level VEV conditions that determine the boundary sheets (or lines) between the unbroken   and broken phases. The vacuum configurations and symmetry breaking patterns are determined from the $\kappa_i$ and as discussed in   section \eqref{loop level eff-analysis}, there is no alternative vacuum configuration found except for $\kappa_i=1\,\left(i=1,2,3\right)$ or $\kappa_i=\delta_{i1}\,\left(i=1,2,3\right)$. 

\begin{figure}[htb]
\centering
\includegraphics[width=0.6\columnwidth]{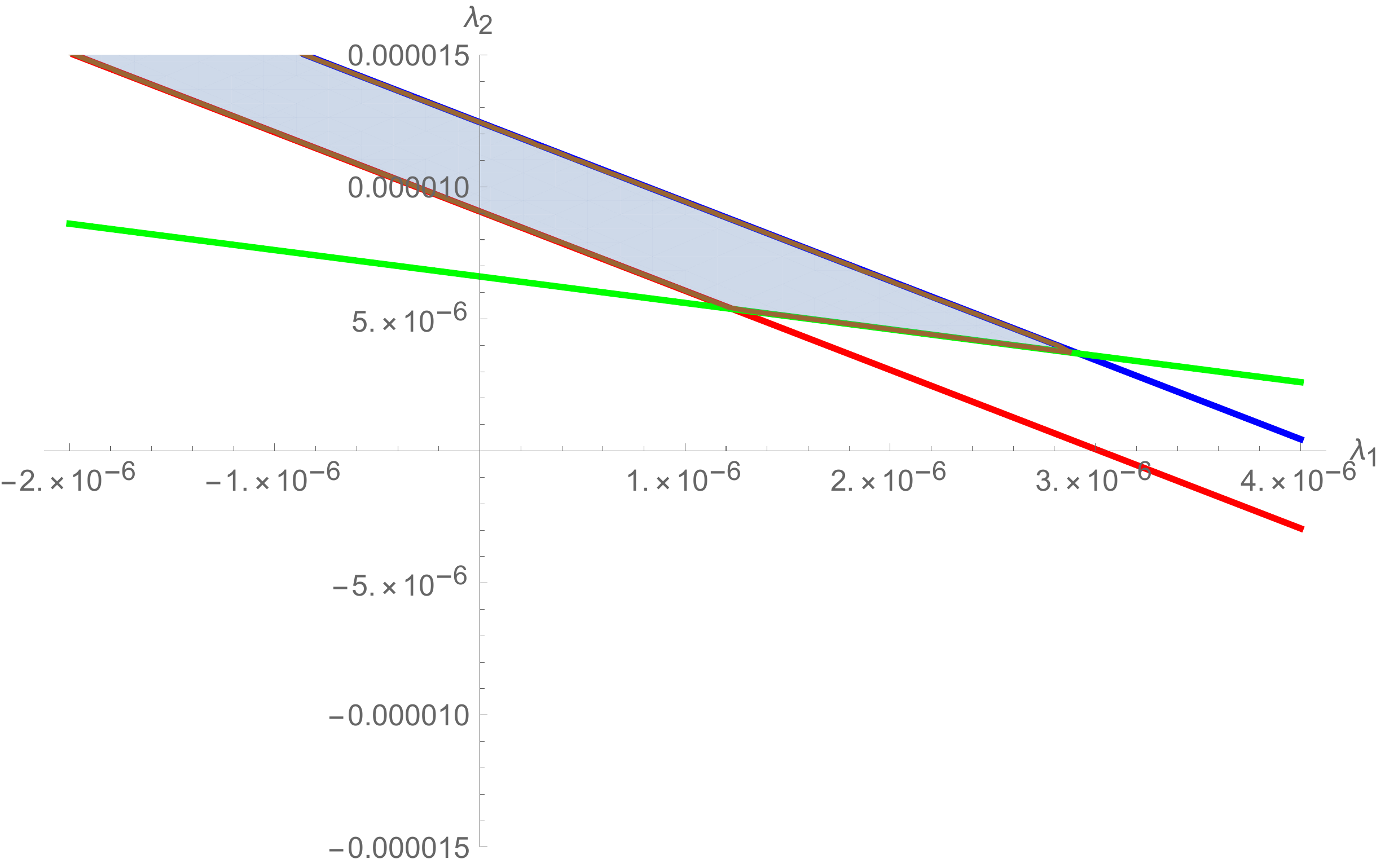}\hspace{0.05\columnwidth}
\caption{In this figure we choose a particular slice $g=0.1$. 
The shaded region represents the broken phases $SU(N_c)\times U(N_s) \to SU(N_c-N_s) \times SU(N_s) \times U(1)$.}
\label{Region Plot_X}
\end{figure}

In the following, we illustrate the case $\kappa_i=1\,\left(i=1,2,3\right)$ as an example. The VEV condition is  
\begin{equation}
\lim_{\kappa_3\rightarrow 1}\frac{\partial V_{eff}^{1loop}}{\partial\rho_i}\bigg|_{\rho_3=\kappa_3M_R\atop \rho_1=\rho_2=M_R}=0\qquad \left(i=1,2,3\right)\,,\label{VEV explicit}
\end{equation}
where the limit is implemented to get rid of the singularity, providing the following constraint 
\begin{equation}
\frac{2}{81} M^3 \left(-\frac{199 g^4}{\pi ^2}+486 \lambda _1+162 \lambda _2\right)=0\label{VEV explicit constraint}\,,
\end{equation}
which corresponds to the blue line in figure~\ref{Region Plot_X}.  All three constraints are equivalent because of the permutation symmetry among $\rho_1, \rho_2, \rho_3$.  
The couplings satisfying the above constraint eq.~\eqref{VEV explicit constraint} are evaluated at the broken scale in the Coleman-Weinberg scheme, whereas the couplings satisfying the constraint eq.~\eqref{RG improved boundary 1} are evaluated at the broken scale in the Minimal-Subtraction scheme. We shall see that the coupling values evaluated in these two  schemes are quite different.

In order to make sure the solutions are at a local minimum  the mass eigenvalues of the Hessian mass matrix 
\begin{equation}
M_{ij}=\lim_{\kappa_3\rightarrow 1}\frac{\partial^2 V_{eff}^{1loop}}{\partial\rho_i\partial\rho_j}\bigg|_{\rho_3=\kappa_3M_R\atop \rho_1=\rho_2=M_R}
\end{equation}
must be non-negative. This  yields 
\begin{equation}
\frac{2}{27}\left(-\frac{145 g^4}{\pi ^2}+486 \lambda _1+162 \lambda _2\right)\geq0,\qquad -\frac{845 g^4}{108 \pi ^2}+12 \lambda _1+12 \lambda _2\geq0\,,\label{vacuum stability conditions explicit}
\end{equation}
which correspond to the red and green lines in figure~\ref{Region Plot_X} respectively.

It is clear that when the RG flows run into the shaded region shown in figure~\ref{Region Plot_X}, the symmetry is broken and we have the symmetry breaking pattern $SU(N_c)\times U(N_s) \to SU(N_c-N_s) \times SU(N_s) \times U(1)$. Comparing figure~\ref{Region Plot_X} with figure~figure~\ref{Region Plot_1}, the shape and the structure of the shaded regions are very similar and consistent, while the coupling solutions are very different in the different schemes. In the Coleman-Weinberg scheme both quartic couplings could be positive and at the same time symmetry breaking is driven by the loop contributions. In the Minimal Subtraction scheme, one of the two quartic couplings is always negative.

\end{document}